\documentclass[singlecolumn,aps,prc,superscriptaddress,nofootinbib,
showpacs,floatfix]{revtex4}
\usepackage{amssymb}

\usepackage{graphicx}

\begin{document}

\title{Low-Energy Thermal Photons from Meson-Meson Bremsstrahlung}
\author{W. Liu and R. Rapp}
\affiliation{Cyclotron Institute and Physics Department, Texas A$\&$M
University, College Station, Texas 77843-3366}
\date{\today}

\begin{abstract}
Within an effective hadronic model including electromagnetic interactions
via a U$_{\rm em}$(1) gauge, we reinvestigate photon Bremsstrahlung from
a hot hadronic gas as expected to be formed in relativistic heavy-ion
collisions at SPS energies.  We calculate photon emission from the
reactions $\pi\pi\to\pi\pi\gamma$ and $\pi K \to\pi K\gamma$ by
an explicit (numerical) evaluation of the multi-dimensional
phase space integral. This, in particular, allows to avoid the
commonly employed soft photon approximation (SPA), as well as to
incorporate final-state thermal enhancement factors.
% during the hadronic stage of the fireball.
Both improvements are shown to result in an appreciable increase
of the photon production rate over previous hadronic calculations.
Upon convolution over a thermal fireball we find an improvement
in the description of recent low transverse-momentum WA98 data at SPS.
The influence of both Landau-Pomeranchuk-Migdal and in-medium effects on
``$\sigma$" and $\rho$-meson exchanges are briefly discussed.
\end{abstract}

\pacs{relativistic heavy-ion collisions, photon spectra, Bremsstrahlung}
\maketitle

%%%%%%%%%%%%%%%%%%%%%%%%%%%%%%%%%%%%%%
\section{Introduction}
%%%%%%%%%%%%%%%%%%%%%%%%%%%%%%%%%%%%%%
A hot deconfined state of matter called quark-gluon plasma (QGP) is
predicted to have existed during the first few microseconds after the
big bang. Ultrarelativistic heavy-ion collisions (URHICs) provide the
only way to recreate such matter for a short moment in the
laboratory. Electromagnetic radiation is expected to be a
unique probe in the study of URHICs, as the (real and virtual) photons,
once produced, decouple from the strongly interacting hot matter.
Since photons are emitted throughout the entire evolution of the
fireball, including QGP and hadronic phases, their spectra
carry valuable information on the various environments of their
creation.

Studies of photon emission from hot and dense matter in heavy-ion
collisions have a long history both theoretically and experimentally,
cf., e.g.,
Refs.~\cite{Alam:1999sc,peitzmann,Arleo:2004gn,rapp04,Stankus:2005eq}
for recent reviews. While scattering in deconfined matter is expected
to mostly manifest itself in the photon spectra at transverse momenta
above $q_t\simeq 2$~GeV, the hadronic stages are likely to dominate the
emission at low
$q_t$~\cite{kapusta91,song93,Golov93,roy96,alam03,TRG04,Haglin:2003sh,Sriva05}.
The latter regime has recently received renewed interest due to
measurements of the WA98 collaboration in central $Pb$-$Pb$ collisions
at the CERN Super Proton Synchrotron (SPS)~\cite{wa98-00,wa98-04}.
Using Hanbury Brown-Twiss (HBT) interferometry methods, a direct
photon signal could be extracted in the range of
$q_t=0.1-0.3$~GeV~\cite{wa98-04}.
These data exhibit a large excess over theoretical
predictions~\cite{TRG04,Sriva05} that have previously shown agreement
with data for the same system (and by the
same experiment)~\cite{wa98-00} at higher momenta, $q_t\ge 1.5$~GeV
(where the more conventional subtraction method enabled the
extraction of a direct photon excess).
In subsequent work~\cite{Sriva05,Tur04,Rapp:2005fy}, the role of
photon Bremsstrahlung from elastic $\pi$-$\pi$ interactions,
which was not included in the previous calculations, has been
re-assessed. It was found that the Bremsstrahlung contribution
notably increases the spectral yield at low $q_t$ by up to 50\% over
the baseline predictions, but still significantly falls short of the
WA98 data~\cite{wa98-04}.

The main objective of the present work is to revisit thermal
Bremsstrahlung from elastic meson-meson interactions.
Based on an effective hadronic Lagrangian, with electromagnetic
interactions implemented via U$_{\rm em}$(1) gauging, we improve the
accuracy of earlier analyses by going beyond the commonly
employed soft photon approximation (SPA)~\cite{ruckl}, and extend the
calculations to the strangeness sector ($\pi$-$K$ scattering). In the
context of soft dileptons, corrections to the SPA have been found to
be sizable and positive~\cite{eggers}.
Our approach furthermore enables the explicit inclusion of thermal
final-state enhancement factors, as well as meson-chemical potentials
which arise in the hadronic evolution of URHICs after hadrochemical
freezeout.
As an indication of the uncertainties in applications to heavy-ion
data, we briefly discuss other meson-meson Bremsstrahlung
sources~\cite{Haglin:2003sh}, as well as schematic estimates
of medium effects in the scalar $\pi$-$\pi$ channel (dropping
$\sigma$ mass) and of the Landau-Pomeranchuk-Migdal (LPM)
effect~\cite{Knoll93,Cleymans93}.
The $\pi$-$\pi$ and $\pi$-$K$ Bremsstrahlung rates are then combined
with earlier hadronic emission calculations~\cite{TRG04} and convoluted
over a thermal fireball model for central $Pb$-$Pb$ collisions at SPS
in order to put the improvements into context with the recent
low-momentum WA98 data~\cite{wa98-04}.

Our article is organized as follows.
In Sec.~\ref{sec_mes} we introduce the chiral effective Lagrangian that
will be used to evaluate amplitudes for
elastic $\pi$-$\pi$ and $\pi$-$K$ interactions, with parameters
fixed to reproduce the corresponding vacuum scattering
data in Born approximation.
In Sec.~\ref{sec_rate}, we discuss our treatment of the
basic kinetic theory expression to calculate Bremsstrahlung
from $\pi$-$\pi$ and $\pi$-$K$ scattering in a thermal bath.
We first construct the pertinent scattering amplitudes by gauging the
hadronic Lagrangian (Sec.~\ref{ssec_ampl}), with special care to
maintain electromagnetic gauge invariance by implementing appropriate
contact interactions (4-point vertices).
We then give a detailed account of the various features that govern
the emission rate (Sec.~\ref{ssec_res}), with emphasis on the
improvements over previous evaluations.
In Sec.~\ref{sec_uncert} we illustrate further uncertainties when
applying low-energy rates to heavy-ion collisions in terms of
schematic estimates of the LPM effect
(Sec.~\ref{ssec_lpm}) and medium modifications of the exchanged
$\rho$ and $\sigma$ mesons (Sec.~\ref{ssec_med}).
In Sec.~\ref{sec_spec}, the soft photon production rate is applied to
calculate photon spectra in central $Pb$-$Pb$ collisions at SPS, which
are supplemented with earlier calculations and compared to
experimental data from WA98.
In Sec.~\ref{sec_coh}, we estimate the coherent photon emission from
initial scattering of protons within both projectile and target nuclei.
Sec.~\ref{sec_sum} contains a summary and discussion.

%%%%%%%%%%%%%%%%%%%%%%%%%%%%%%%%%%%%%%%%%%%%%%%%%%%%%%%%%%%%%%
\section{Elastic Meson-Meson Scattering}
\label{sec_mes}
\begin{figure}[th]
\includegraphics[height=1.5in,width=5.0in]{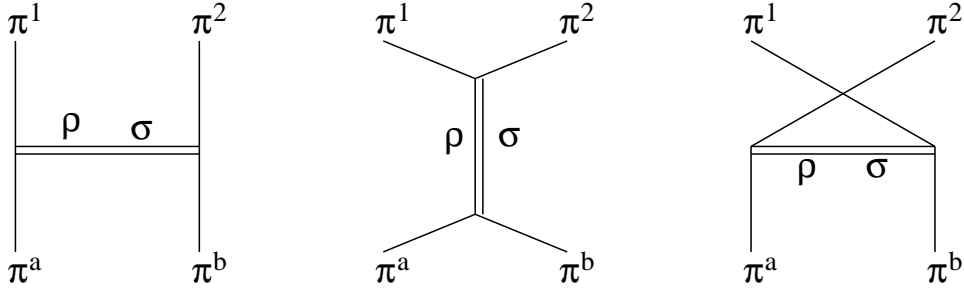}
\caption{Diagrams for $\pi^a\pi^b\to\pi^1\pi^2$ scattering.}
\label{fig_dia1}
\end{figure}

\begin{figure}[ht]
\centerline{
\includegraphics[height=8.0cm,width=7.5cm,angle=-90]{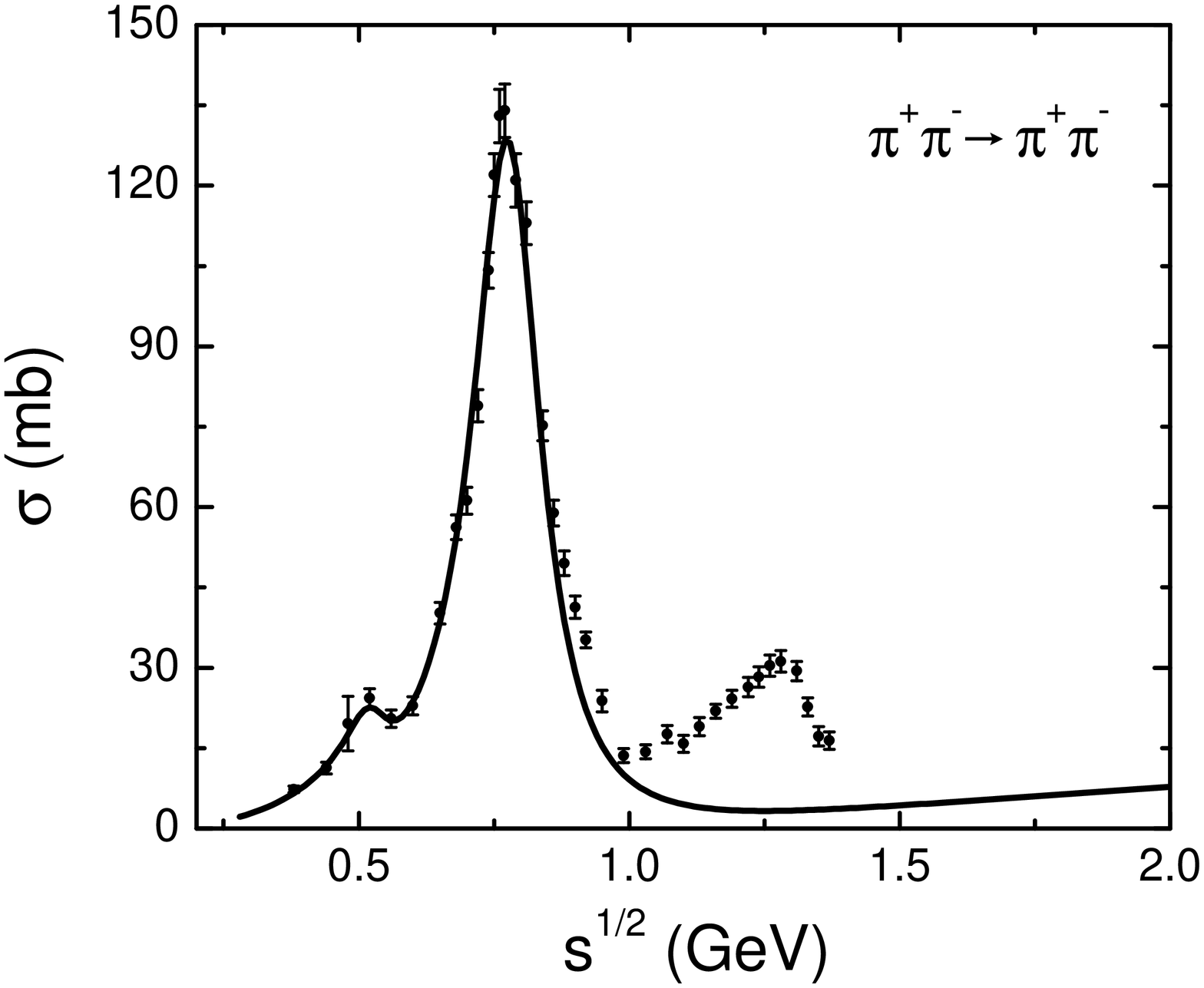}
\hspace{0.8cm}
\includegraphics[height=8.0cm,width=7.5cm,angle=-90]{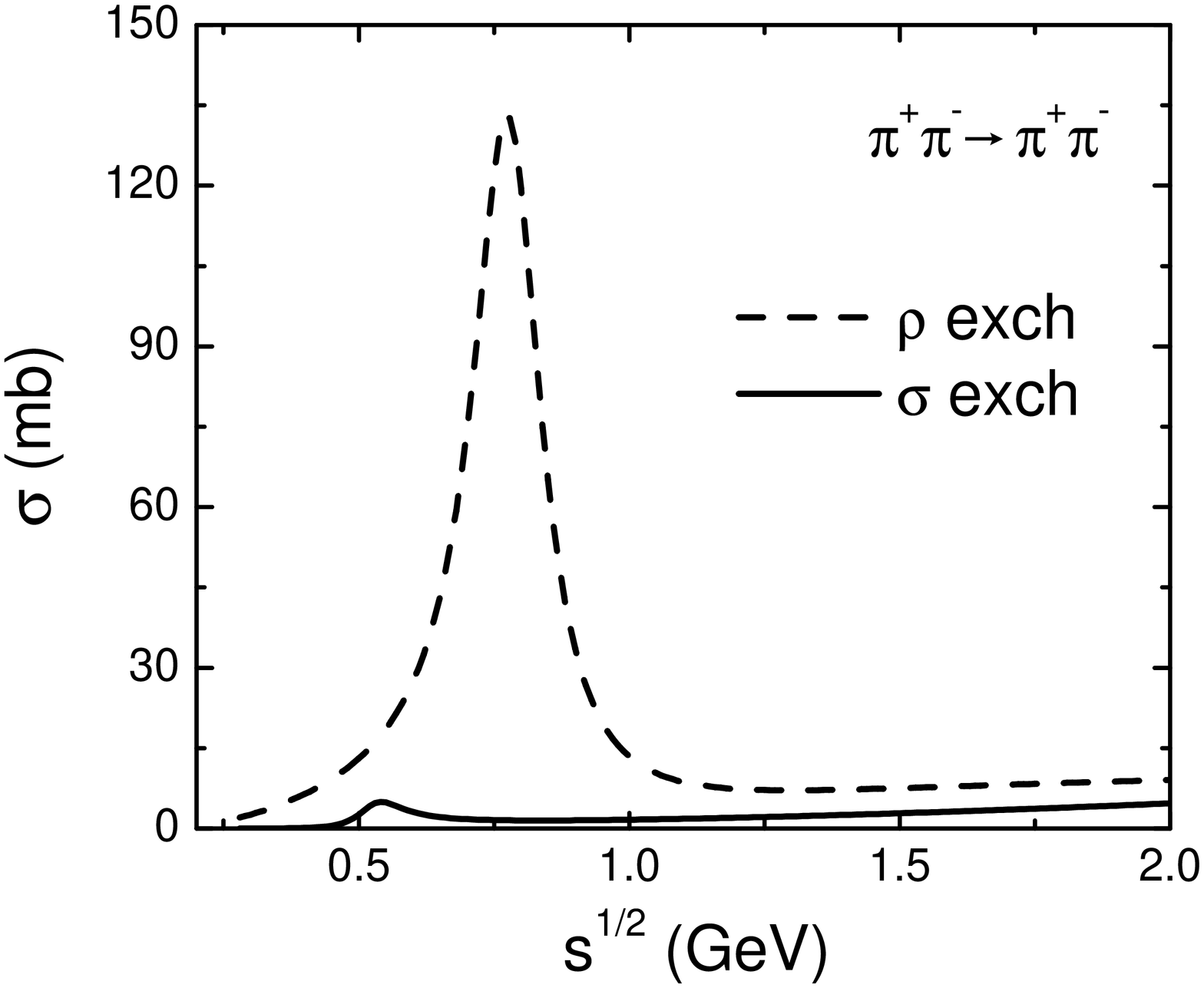}}
\caption{Left panel: data on the elastic $\pi^+\pi^-\to\pi^+\pi^-$
cross section~\cite{pipi-data} compared to a fit
using a meson-exchange model in Born approximation.
Right panel: decomposition of the cross section into
contributions from $\sigma$- and $\rho$-exchange.}
\label{fig_sigpp}
\end{figure}

%%%%%%%%%%%%%%%%%%%%%%%%%%%%%%%%%%%%%%%%%%%%%%%%%%%%%%%%%%%%%%
In this work we focus on photon Bremsstrahlung induced by elastic
interactions of the lightest constituents in a hadronic gas, i.e.,
pion-pion ($\pi$-$\pi$) and pion-kaon ($\pi$-$K$) scattering (the
latter will turn out to produce a factor of 4-5 less low-energy photons
than the former; other contributions will be briefly discussed at the
end of Sec.~\ref{ssec_res}). We first recall the underlying
model~\cite{haglin93,roy96} that will allow us reproduce vacuum
scattering data and thus
form the basis to evaluate the pertinent Bremsstrahlung processes,
$\pi^a\pi^b\to\pi^1\pi^2\gamma$ and $\pi^a K^b\to\pi^1 K^2\gamma$.
%in the hadronic phase of the fireball formed in URHICs.
Starting point is a chiral Lagrangian for the pseudoscalar
fields ($\pi$ and $K$)~\cite{changhui} where the low-lying vector
mesons, $\rho$ and $K^*$, are introduced via a covariant derivative
leading to
\begin{eqnarray}
{\cal L}&=&g_{\sigma\pi\pi}\sigma\partial_\mu\vec\pi\cdot\partial^\mu\vec\pi
+g_{\rho\pi\pi}\vec\rho^\mu
\cdot(\vec\pi\times\partial_\mu\vec\pi)
+ig_{\rho KK}(\bar K\vec\tau\partial_\mu K-
\partial_\mu\bar K\vec\tau K)\cdot\vec\rho^\mu\nonumber\\
&&-ig_{\pi KK^*}(\bar K\vec\tau K^{*\mu}\cdot\partial_\mu\vec\pi
-\partial_\mu \bar K\vec\tau K^{*\mu}\cdot\vec\pi)+{\rm h.c.} \ .
\end{eqnarray}
Since we will restrict ourselves to the Born approximation, an
additional (chirally invariant) $\sigma$-$\pi$-$\pi$ interaction
vertex with a low-mass ``$\sigma$"-meson~\cite{donoghue} has been added
to adequately account for $S$-wave $\pi$-$\pi$ scattering near threshold.
The corresponding Feynman diagrams, comprising $s$, $t$ and $u$ channel
exchanges, are depicted in Fig.~\ref{fig_dia1}.
The pertinent amplitudes for elastic $\pi^+$-$\pi^-$ scattering
in $t$- and $s$-channel are given by
\begin{eqnarray}
{\cal M}_t(p_a,p_b,p_1,p_2)&=&
\frac{-4g^2_{\sigma\pi\pi}F^2(t)(p_a\cdot p_1)(p_b\cdot p_2)}
{t-m^2_\sigma+im_\sigma\Gamma_\sigma}
+\frac{-g^2_{\rho\pi\pi}F^2(t)
(p_a\cdot p_b+p_a\cdot p_2+p_b\cdot p_1+p_1\cdot p_2)}
{t-m^2_\rho+im_\rho\Gamma_\rho},
\nonumber\\
{\cal M}_s(p_a,p_b,p_1,p_2)&=&
\frac{-4g^2_{\sigma\pi\pi}(p_a\cdot p_b)(p_1\cdot p_2)}
{s-m^2_\sigma+im_\sigma\Gamma_\sigma}
+\frac{g^2_{\rho\pi\pi}(p_a\cdot p_1-p_a\cdot p_2+p_b\cdot p_2-p_b\cdot p_1)}
{s-m^2_\rho+im_\rho\Gamma_\rho}.
\end{eqnarray}
To account for the finite size of the hadronic vertices, we introduce a
momentum-transfer damping monopole formfactor, $F(q^2)$, for the
$t$ and $u$ channels~\cite{haglin93,roy96},
\begin{equation}
F(q^2)=\frac{m^2_\alpha-m^2_\pi}{m^2_\alpha-q^2} \ ,
\end{equation}
where $m_\alpha$ is the mass of the exchanged meson and $q$ the
four-momentum transfer (for simplicity, we neglect formfactors for
$s$-channel polegraphs).
The differential cross section for $\pi^+\pi^-\to\pi^+\pi^-$ then
follows as
\begin{eqnarray}
\frac{d\sigma}{dt}=\frac{|{\cal M}_t+{\cal M}_s|^2}{64\pi s p^2_{in}}  \ ,
\label{dcross}
\end{eqnarray}
with $p_{in}=\sqrt{(s-4m_\pi^2)}/2$ the initial pion three-momentum in
the center-of-mass (cm) frame.
Fixing the parameters in the non-strange sector as in
Refs.~\cite{haglin93,roy96}, $m_\sigma=0.525$~GeV,
$\Gamma_\sigma=0.1$~GeV, $g_{\sigma\pi\pi}m_\sigma=1.85$,
$m_\rho=0.775$~GeV, $\Gamma_\rho=0.155$~GeV and $g_{\rho\pi\pi}=6.15$,
provides a fair fit to the total $\pi\pi$ cross section~\cite{pipi-data}
up to a cm energy of $\sim$1~GeV, cf.~left panel of Fig.~\ref{fig_sigpp}.
We neglect the contribution from the $f_2(1270)$ tensor meson as
its thermal density is a factor of $\sim$25 smaller than $\rho$-density.
In the right panel, we display the decomposition into $\sigma$- and
$\rho$-meson exchanges which identifies the $s$-channel $\rho$-polegraph
as the prevalent contribution to elastic $\pi^+\pi^-$ scattering,
while the one from $\sigma$-exchange is small.

\begin{figure}[!th]
\includegraphics[height=9cm,width=8cm,angle=-90]{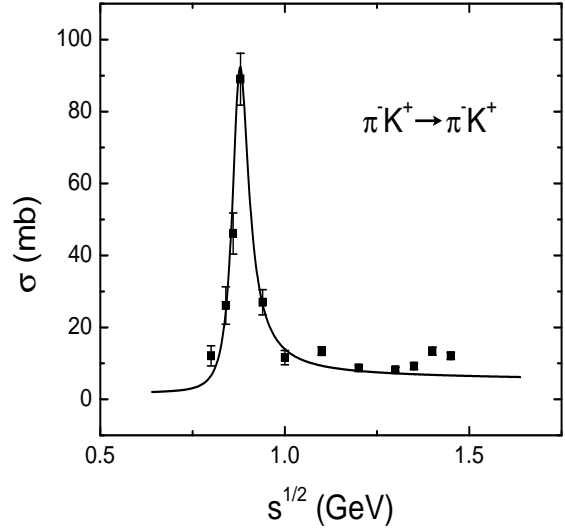}
\caption{Total elastic $\pi^-K^+\to\pi^-K$ cross section as calculated
in a meson-exchange model in Born approximation compared to
experimental data from Ref.~\cite{firestone}.}
\label{fig_sigpk}
\end{figure}

Next, we turn to $\pi$-$K$ scattering, Bremsstrahlung off which could
be significant due to both the kaon's relatively large abundance in
the hadronic fireball and its resonance cross
section via the $K^*$ resonance.
The relevant Feynman diagrams follow from Fig.~\ref{fig_dia1} by
replacing one of each in- and outgoing pion by a kaon, and the $\rho$
by $K^*$ (no $\sigma$-exchange), resulting in sixteen processes with
charged particles in the initial and/or final state.  The corresponding
amplitudes for elastic $\pi^-$-$K^+$ scattering are given by
\begin{eqnarray}
{\cal M}_t(p_a,p_b,p_1,p_2)&=&\frac{-g_{\rho\pi\pi}g_{\rho KK}F^2(t)
(p_a\cdot p_b+p_a\cdot p_2+p_b\cdot p_1+p_1\cdot p_2)}
{t-m^2_\rho+im_\rho\Gamma_\rho}\nonumber\\
{\cal M}_s(p_a,p_b,p_1,p_2)&=&\frac{2g^2_{\pi KK^*}}
{s-m^2_{K^*}+im_{K^*}\Gamma_{K^*}}\left[
(p_a\cdot p_1-p_a\cdot p_2+p_b\cdot p_2-p_b\cdot p_1)\right.\nonumber\\
&&\left.-\frac{(m^2_\pi-m^2_K)
(p_a\cdot p_1-p_a\cdot p_2+p_b\cdot p_1-p_b\cdot p_2)}{m^2_{K^*}}\right] .
\end{eqnarray}
As before, the hadronic $t$- and $u$-channel vertices are augmented by
a formfactor,
\begin{equation}
F(q^2)=\frac{\Lambda^2-m^2_\alpha}{\Lambda^2_\alpha-q^2} \ ,
\end{equation}
which is of the same form as above but with a slightly different
parameter dependence which facilitates a better fit to the free
scattering data.
Applying Eq.~(\ref{dcross}) and fixing $\Lambda=1.5$~GeV,
$\Gamma_{K^*}=0.051$~GeV, $g_{\rho KK}=g_{\pi KK^*}=3.55$, and
$m_{K^*}=0.89$~GeV, the empirical cross section for elastic
$\pi^-K^+\to\pi^-K^+$ scattering~\cite{firestone} is reasonably well
described, cf.~Fig.~\ref{fig_sigpk}.

%%%%%%%%%%%%%%%%%%%%%%%%%%%%%%%%%%%%%%%%%%%%%%%%%%%%%%%%%
\section{Thermal Photon Emission Rate}
\label{sec_rate}
%%%%%%%%%%%%%%%%%%%%%%%%%%%%%%%%%%%%%%%%%%%%%%%%%%%%%%%%%
In kinetic theory, and to leading order in the e.m.~coupling constant
$\alpha_{em}=1/137$, the thermal emission rate (per unit four-volume)
of photons of energy $q_0=q$ for
a Bremsstrahlung process of type $a+b\to 1+2+\gamma$ can be cast
into the form
\begin{eqnarray}
q_0\frac{dR_\gamma}{d^3q}
&=&N\int\frac{d^3p_a}{2E_a(2\pi)^3}\frac{d^3p_b}{2E_b(2\pi)^3}
\frac{d^3p_1}{2E_1(2\pi)^3}\frac{d^3p_2}{2E_2(2\pi)^3}
\nonumber\\
&&\times(2\pi)^4\delta^4(p_a+p_b-p_1-p_2-q)|{\cal M}_\gamma|^2
\frac{f(E_a)f(E_b)[1\pm f(E_1)][1\pm f(E_2)]}{2(2\pi)^3} \ ,
\label{rate0}
\end{eqnarray}
where ${\cal M}_\gamma$ is the corresponding scattering amplitude,
$f(E_i)=1/({\rm e}^{(E_i-\mu_i)/T}\pm 1)\equiv f_i$ are
Fermi-Dirac or Bose-Einstein distribution functions
($\mu_i$: chemical potential of hadron $i$), and
the overall degeneracy factor $N$ depends on the specific process.
As elaborated in Appendix~\ref{app_integral}, we rewrite the emission
rate without further approximation as
\begin{eqnarray}
q_0\frac{dR_\gamma}{d^3q}&=&\frac{N}{16(2\pi)^{11}}
\int |{\bf p}_a|dE_ad\phi_a\int|{\bf p}_b|dE_b{\rm sin\theta_b}
d{\rm \theta_b}d{\rm \phi_b}\int|
{\bf p}_1|dE_1{\rm sin\theta_1}d{\rm \theta_1}
d{\rm \phi_1}f_af_b[1+f_1][1+f_2]\frac{|{\cal M}_\gamma|^2}{\cal A}
\label{rate}
\end{eqnarray}
where ${\cal A}$ is given by
\begin{eqnarray}
{\cal A}=|\varphi'({\rm cos\theta^r_a})|
\end{eqnarray}
and ${\rm cos\theta^r_a}$ is the root of the function
$\varphi({\rm cos\theta_a})$,
\begin{eqnarray}
\varphi({\rm cos\theta_a})
&=&(2|{\bf p}_a||{\bf p}_b|{\rm sin\theta_bcos\phi_acos\phi_b}
+2|{\bf p}_a||{\bf p}_b|{\rm sin\theta_bsin\phi_asin\phi_b}
-2|{\bf p}_a||{\bf p}_1|{\rm sin\theta_1cos\phi_acos\phi_1}
\nonumber\\
&&-2|{\bf p}_a||{\bf p}_1|{\rm sin\theta_1sin\phi_asin\phi_1})
\sqrt{1-{\rm cos}^2{\rm \theta_a}}
\nonumber\\
&&+(2|{\bf p}_a||{\bf p}_b|{\rm cos\theta_b}-2|{\bf p}_a||{\bf p}_1|
{\rm cos\theta_1} -2|{\bf p}_a||{\bf q}|){\rm cos\theta_a}
\nonumber\\
&&+|{\bf p}_a|^2+({\bf p}_b-{\bf p}_1-{\bf q})^2
-(E_a+E_b-E_1-q_0)^2+m^2_2 \ =0 \ .
\end{eqnarray}
The main point of the present work is that the numerical
evaluation of Eq.~(\ref{rate}) goes beyond previous
treatments~\cite{haglin93,roy96} by including the final state
enhancement factors, $(1+f_1)\times(1+f_2)$\footnote{The relevance
of the final-state enhancement factors in hadronic phase of URHICs is
further augmented by the build-up of substantial pion- and
kaon-chemical potentials after chemical freezeout~\cite{Rapp:2002}.},
and by evaluating ${\cal M}_\gamma$ beyond the soft
photon approximation (SPA). The latter point is detailed in the
following Section, where a summation over all possible processes for
thermal photon Bremsstrahlung from $\pi$-$\pi$ and $\pi$-$K$ scattering
is performed.

%%%%%%%%%%%%%%%%%%%%%%%%%%%%%%%%%%%%%%%%%%%%%%%%%%%%
\subsection{Amplitudes for Bremsstrahlung}
\label{ssec_ampl}
%%%%%%%%%%%%%%%%%%%%%%%%%%%%%%%%%%%%%%%%%%%%%%%%%%%%
The photon coupling to pseudoscalar and vector mesons
can be implemented via a ${\rm U_{em}(1)}$ gauge~\cite{song93},
rendering the interaction Lagrangian with electromagnetism
as
\begin{eqnarray}
{\cal L}&=&eA^\mu(\partial_\mu\vec\pi\times\vec\pi)_3
+2eg_{\sigma\pi\pi}\sigma A^\mu(\partial_\mu\vec\pi\times\vec\pi)_3-eg_{\rho\pi\pi}
 A_\mu[\vec\pi\times(\vec\pi\times\vec\rho^\mu)]_3\nonumber\\
&&+e\{A^{\mu}
(\partial_{\mu}\vec\rho^{\nu}\times\vec{\rho}_{\nu})_3
+[(\partial_{\mu}A^\nu \vec\rho_{\nu}-A^{\nu}
\partial_{\mu}\vec\rho_{\nu})\times\vec\rho^\mu]_3+[\vec\rho^\mu\times(A^{\nu}
\partial_{\mu}\vec\rho_\nu-\partial_{\mu}A^{\nu}\vec\rho_{\nu})]_3\}\nonumber\\
&&+ie[A^{\mu}(\bar K^{*\nu}Q\partial_\mu K^*_\nu-\partial_\mu\bar K^{*\nu}Q K^*_\nu)
+\bar K^{*\mu}Q(\partial_{\mu}A^\nu K^*_\nu -A^{\nu}\partial_{\mu}K^*_\nu)
+(A^\nu\partial_\mu\bar K^*_\nu-\partial A^\nu\bar K^*_\nu)QK^{*\mu}]\nonumber\\&&+ieA^\mu(\bar KQ\partial_\mu K-\partial_\mu\bar KQK)
-eg_{\pi KK^*}A^\mu[\bar K^*(2\vec\tau Q-Q\vec\tau)K
+\bar K(2Q\vec\tau-\vec\tau Q)K^*]\cdot\vec\pi\nonumber\\
&&+eg_{\rho KK}A^\mu\bar K(\vec\tau Q+Q\vec\tau)K\cdot\vec\rho_\mu,
\end{eqnarray}
where $A^\mu$ is the electromagnetic field, $Q$=diag(1, 0) the
charge operator, and the subscript ``3" denotes the third component
of isospin.

First, we consider the process for Bremsstrahlung of a photon via
$\pi^a\pi^b\to\pi^1\pi^2\gamma$ scattering, based on
the diagrams depicted in Fig.~\ref{fig_dia1}. In charge basis, this
amounts to the following seven processes:
$\pi^+\pi^-\to\pi^+\pi^-(\pi^0\pi^0)$, $\pi^+\pi^0\to\pi^+\pi^0$,
$\pi^-\pi^0\to\pi^-\pi^0$, $\pi^0\pi^0\to\pi^+\pi^-$,
$\pi^+\pi^+\to\pi^+\pi^+$ and $\pi^-\pi^-\to\pi^-\pi^-$,
with a photon attached to each external or internal charged particle.
E.g., the amplitude for the process $\pi^+\pi^-\to\pi^+\pi^-\gamma$ with
a photon attached to a charged external pion in all possible ways
reads
\begin{eqnarray}
{\cal M}^\mu&=&eJ^\mu_a[{\cal M}_t(p_a-q,p_b,p_1,p_2)
+{\cal M}_s(p_a-q,p_b,p_1,p_2)]\nonumber\\
&&+eJ^\mu_b[{\cal M}_t(p_a,p_b-q,p_1,p_2)
+{\cal M}_s(p_a,p_b-q,p_1,p_2)]
\nonumber\\
&&+eJ^\mu_1[{\cal M}_t(p_a,p_b,p_1+q,p_2)
+{\cal M}_s(p_a,p_b,p_1+q,p_2)]
\nonumber\\
&&+eJ^\mu_2[{\cal M}_t(p_a,p_b,p_1,p_2+q)
+{\cal M}_s(p_a,p_b,p_1,p_2+q)] \ ,
\label{gam-ampl}
\end{eqnarray}
where ${\cal M}_s$ and ${\cal M}_t$ are the amplitudes quoted in
Sec.~\ref{sec_mes}. In the above,
\begin{eqnarray}
J^\mu_{a,b}=\frac{-Q_{a,b}(2p_{a,b}-q)^\mu}{2p_{a,b}\cdot q},~
J^\mu_{1,2}=\frac{Q_{1,2}(2p_{1,2}+q)^\mu}{2p_{1,2}\cdot q}
\end{eqnarray}
are the electromagnetic currents with $Q_i$ the charge of the pion
in units of the proton charge.  Electromagnetic gauge invariance
furthermore requires the incorporation of contact diagrams with the
photon attached to each proper vertex.
In Appendix~\ref{app_gauge} we explicitly construct the
pertinent amplitudes so that the total satisfies
$q_\mu\cdot{\cal M}^\mu=0$.
For the processes with four-point vertices one arrives at
\begin{eqnarray}
{\cal M}^\mu_c&=&\frac{4eg^2_{\sigma\pi\pi}F[(p_b-p_2)^2](p_a+p_1)^\mu p_b\cdot p_2}
{(p_b-p_2)^2-m^2_{\sigma}+im_\sigma\Gamma_\sigma}
-\frac{4eg^2_{\sigma\pi\pi}F[(p_a-p_1)^2]p_a\cdot p_1(p_b+p_2)^\mu}
{(p_a-p_1)^2-m^2_{\sigma}+im_\sigma\Gamma_\sigma}\nonumber\\
&&-\frac{4eg^2_{\sigma\pi\pi}(p_a-p_b)^\mu p_1\cdot p_2}
{(p_1+p_2)^2-m^2_{\sigma}+im_\sigma\Gamma_\sigma}
-\frac{4eg^2_{\sigma\pi\pi}p_a\cdot p_b(p_1-p_2)^\mu}
{(p_a+p_b)^2-m^2_{\sigma}+im_\sigma\Gamma_\sigma}\nonumber\\
&&-\frac{2eg^2_{\rho\pi\pi}F[(p_a-p_1)^2](p_a+p_1)^\mu}
{(p_a-p_1)^2-m^2_{\rho}+im_\rho\Gamma_\rho}
+\frac{2eg^2_{\rho\pi\pi}F[(p_b-p_2)^2](p_b+p_2)^\mu}
{(p_b-p_2)^2-m^2_{\rho}+im_\rho\Gamma_\rho}\nonumber\\
&&-\frac{2eg^2_{\rho\pi\pi}(p_a-p_b)^\mu}{(p_a+p_b)^2-m^2_{\rho}+im_\rho\Gamma_\rho}
-\frac{2eg^2_{\rho\pi\pi}(p_1-p_2)^\mu}{(p_1+p_2)^2-m^2_{\rho}
+im_\rho\Gamma_\rho} \ .
\end{eqnarray}

The same procedure of deriving the Bremsstrahlung amplitude, leading
to Eqs.~(\ref{gam-ampl}), can be applied to $\pi$-$K$ scattering.
The additional contact terms necessary to maintain gauge invariance
are given by
\begin{eqnarray}
{\cal M}^\mu_c&=&-\frac{2eg_{\rho\pi\pi}g_{\rho KK}F[(p_a-p_1)^2](p_a+p_1)^\mu}
{(p_a-p_1)^2-m^2_{\rho}+im_\rho\Gamma_\rho}
+\frac{2eg_{\rho\pi\pi}g_{\rho KK}F[(p_b-p_2)^2](p_b+p_2)^\mu}
{(p_b-p_2)^2-m^2_{\rho}+im_\rho\Gamma_\rho}\nonumber\\
&&-\frac{4eg^2_{\pi KK^*}[(p_a-p_b)^\mu-(m^2_a-m^2_b)(p_a+p_b)^\mu/m^2_{K^*}]}
{(p_a+p_b)^2-m^2_{K^*}+im_{K^*}\Gamma_{K^*}}\nonumber\\
&&-\frac{4eg^2_{\pi KK^*}[(p_1-p_2)^\mu-(m^2_1-m^2_2)(p_1+p_2)^\mu/m^2_{K^*}]}
{(p_1+p_2)^2-m^2_{K^*}+im_{K^*}\Gamma_{K^*}} \ ,
\end{eqnarray}
which completes the set of processes for $\pi K \to \pi K \gamma$.

%%%%%%%%%%%%%%%%%%%%%%%%%%%%%%%%%%%%%%%%%%%%%%%%%%%%
\subsection{Soft Photon Emission Rates}
\label{ssec_res}
\begin{figure}[!t]
\centerline{
\includegraphics[height=10.0cm,width=8.0cm,angle=-90]{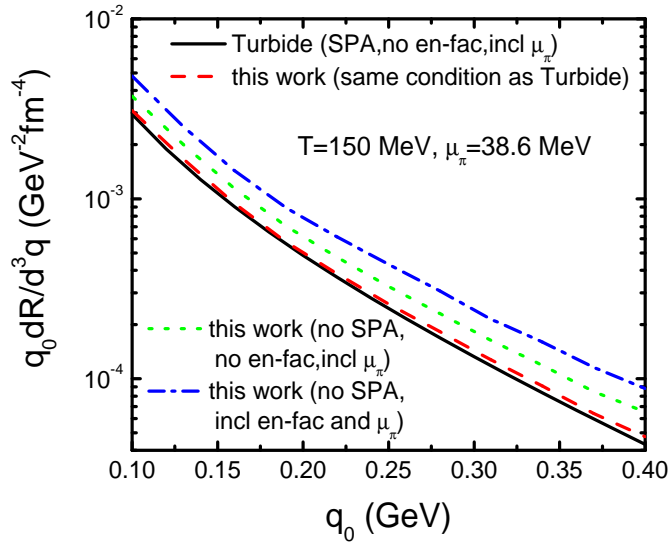}}
\caption{Comparison of various approximations to the thermal photon
rate for $\pi\pi\to\pi\pi\gamma$ reactions as a function
of photon energy.}
\label{fig_comp}
\end{figure}

%%%%%%%%%%%%%%%%%%%%%%%%%%%%%%%%%%%%%%%%%%%%%%%%%%%%
With the amplitudes as specified in the previous section we now
proceed to the results of the numerical integration of the thermal
photon production rate, Eq.~(\ref{rate}). In Fig.~\ref{fig_comp} we
compare our new results for Bremsstrahlung off $\pi$-$\pi$ scattering
to earlier calculations. With the application to URHICs in mind, we
choose medium conditions representing $Pb$-$Pb$ collisions at full SPS
energy, roughly half way through
the hadronic evolution, corresponding to a temperature of
$T=150$~MeV and a pion chemical potential of
$\mu_\pi\simeq 40$~MeV~\cite{RW99,TRG04}.
When working in SPA and neglecting
final-state Bose enhancement factors for the outgoing pions, our
results closely coincide with computations by Turbide
et al.~\cite{Tur04,Rapp:2005fy} (dashed vs. solid line in
Fig.~\ref{fig_comp}).
When the SPA is relaxed, the rate is found to increase by about 25\%
at low energies ($q_0=0.1$~GeV) and close to 40\% at higher energies
($q_0\simeq0.4$~GeV), see the short-dashed line in Fig.~\ref{fig_comp}.
Finally, when implementing the final-state Bose enhancement, the rate
rises another $\sim$30\%. Thus, the total increase of soft photon
emission over previous computations amounts to 60-100\%, which is quite
appreciable and the main result of our paper.

\begin{figure}[!tb]
\centerline{
\includegraphics[height=8.0cm,width=7.5cm,angle=-90]{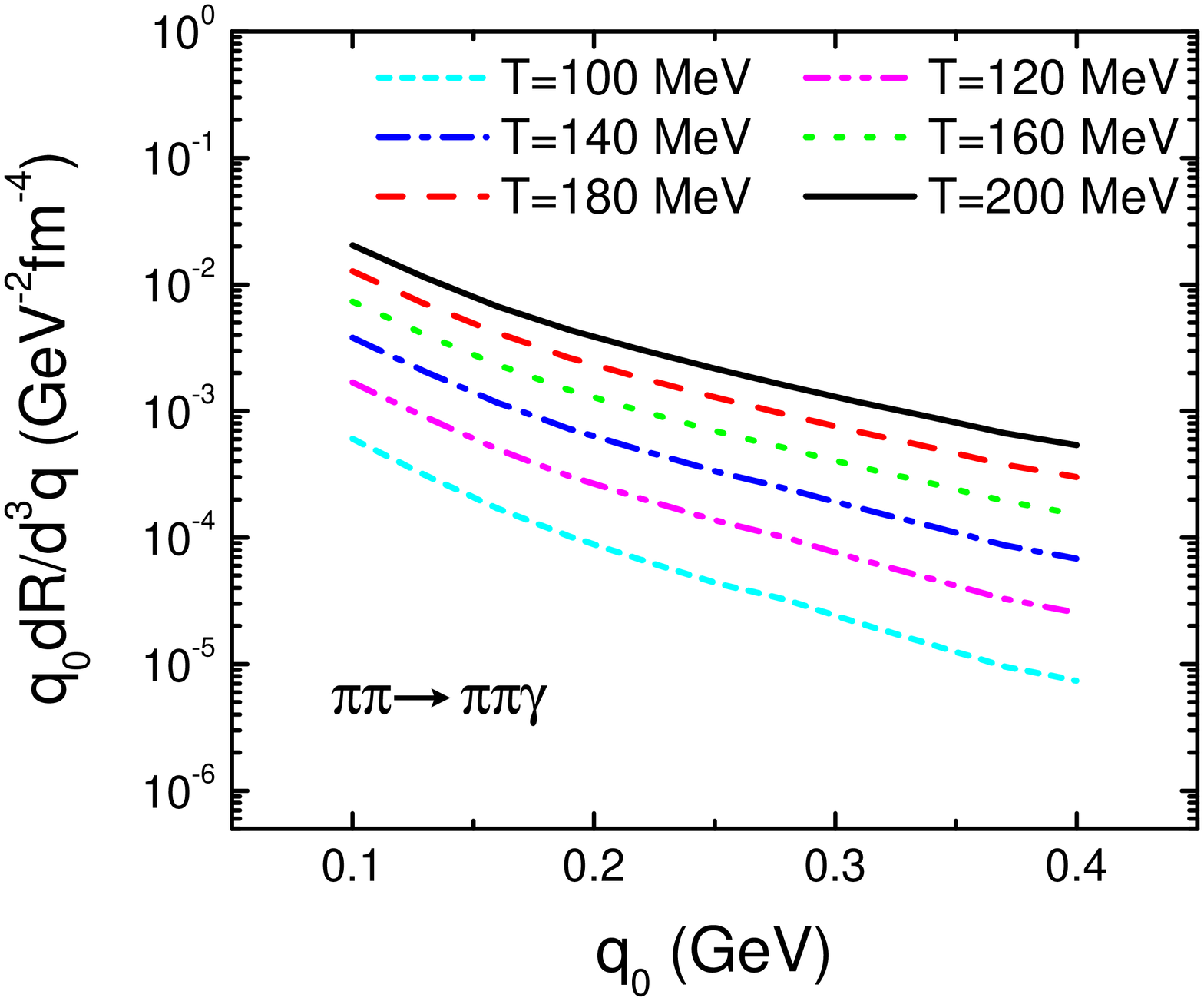}
\hspace{0.8cm}
\includegraphics[height=8.0cm,width=7.5cm,angle=-90]{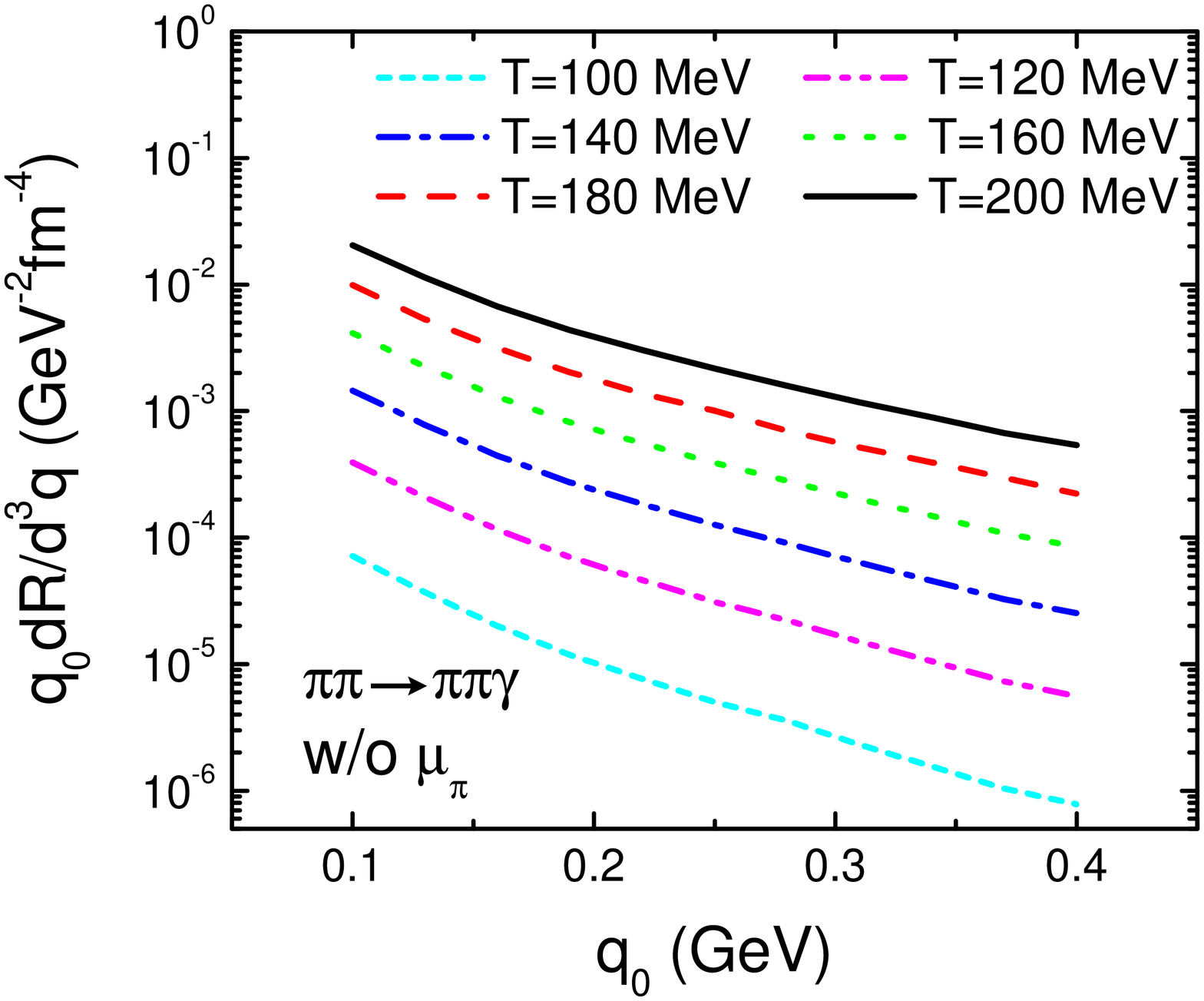}}
\caption{Thermal photon emission rate from Bremsstrahlung via
$\pi\pi\to\pi\pi\gamma$ reactions as a function of photon energy at
different temperatures.
The left panel includes pion chemical potentials appropriate for
$Pb$(158~AGeV)-$Pb$ collisions while in the right panel $\mu_\pi=0$.
}
\label{fig_ratepi}
\end{figure}
In Fig.~\ref{fig_ratepi} we display the temperature evolution of the
Bremsstrahlung rate from $\pi$-$\pi$ scattering. The left panel
includes pion-chemical potentials as estimated in Ref.~\cite{RW99}
to preserve the pion (and kaon) number in the hadronic evolution of
$Pb$-$Pb$ collisions after chemical freezeout, parametrized by a linear
increase with temperature ($T$ in units of GeV),
\begin{eqnarray}
\mu_\pi=\frac{0.2-T}{1.05}~{\rm GeV}\ ,
\quad \mu_K=1.9~(0.175-T)~{\rm GeV}
\label{chem}
\end{eqnarray}
(a small nonvanishing value of $\mu_\pi$ at chemical freezeout has
been introduced to reproduce the measured pion-to-baryon ratio at full
SPS energy; this is due to the somewhat limited number of resonances
included in the equation of state in Ref.~\cite{RW99} which affects the
pion yield via feeddown; it slightly overestimates (underestimates)
the direct pion density at large (small) temperature, e.g.,
${\rm e}^{\mu_\pi/T}=1.11$ at $T=0.18$~GeV). When reducing
the temperature from $T=0.18$~GeV to 0.11~GeV, the decrease in rate
is close to a factor of 10 at $q_0=0.1$~GeV but becomes a factor of
$\sim$15 at $q_0=0.4$~GeV, reflecting the larger slope at smaller $T$.
This also entails an increasing weight
of photon emission at lower energies toward later stages in the
fireball evolution (note that increase in the hadronic fireball volume
between chemical ($T_c=0.175$~GeV) and thermal freezeout
($T\simeq0.110$~GeV) is about a factor of $\sim$6, implying that the
yield of Bremsstrahlung photons
is rather sensitive to the late stages of the fireball lifetime).
In the right panel of Fig.~\ref{fig_ratepi}, the pion-chemical
potentials are set to zero: the decrease compared to the $\mu_\pi\ne 0$
results amounts to slightly more than a fugacity factor squared,
${\rm e}^{2\mu_\pi/T}$ (as one would naively expect for $\pi\pi$
annihilation), mostly due to the final-state enhancement factors.
Finally, a comparison to the hadronic rate calculations of
Ref.~\cite{TRG04} reveals that $\pi$-$\pi$ Bremsstrahlung as computed
here exceeds previously dominant baryonic contributions
by close to a factor of 2 at $q_0=0.2$~GeV, and is smaller by a factor
of 2 at $q_0=0.4$~GeV.

\begin{figure}[!t]
\centerline{
\includegraphics[height=8.0cm,width=7.5cm,angle=-90]{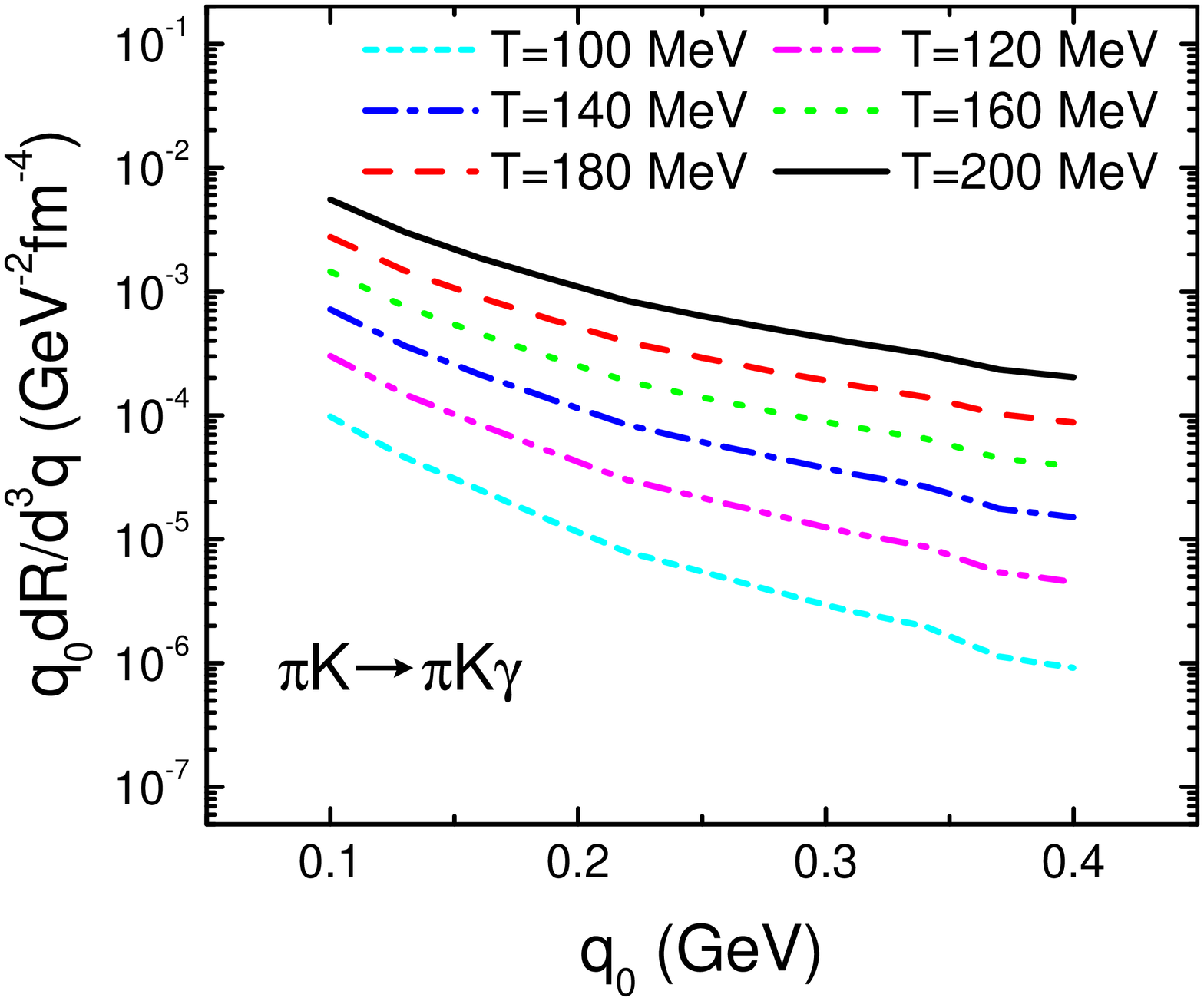}
\hspace{0.8cm}
\includegraphics[height=8.0cm,width=7.5cm,angle=-90]{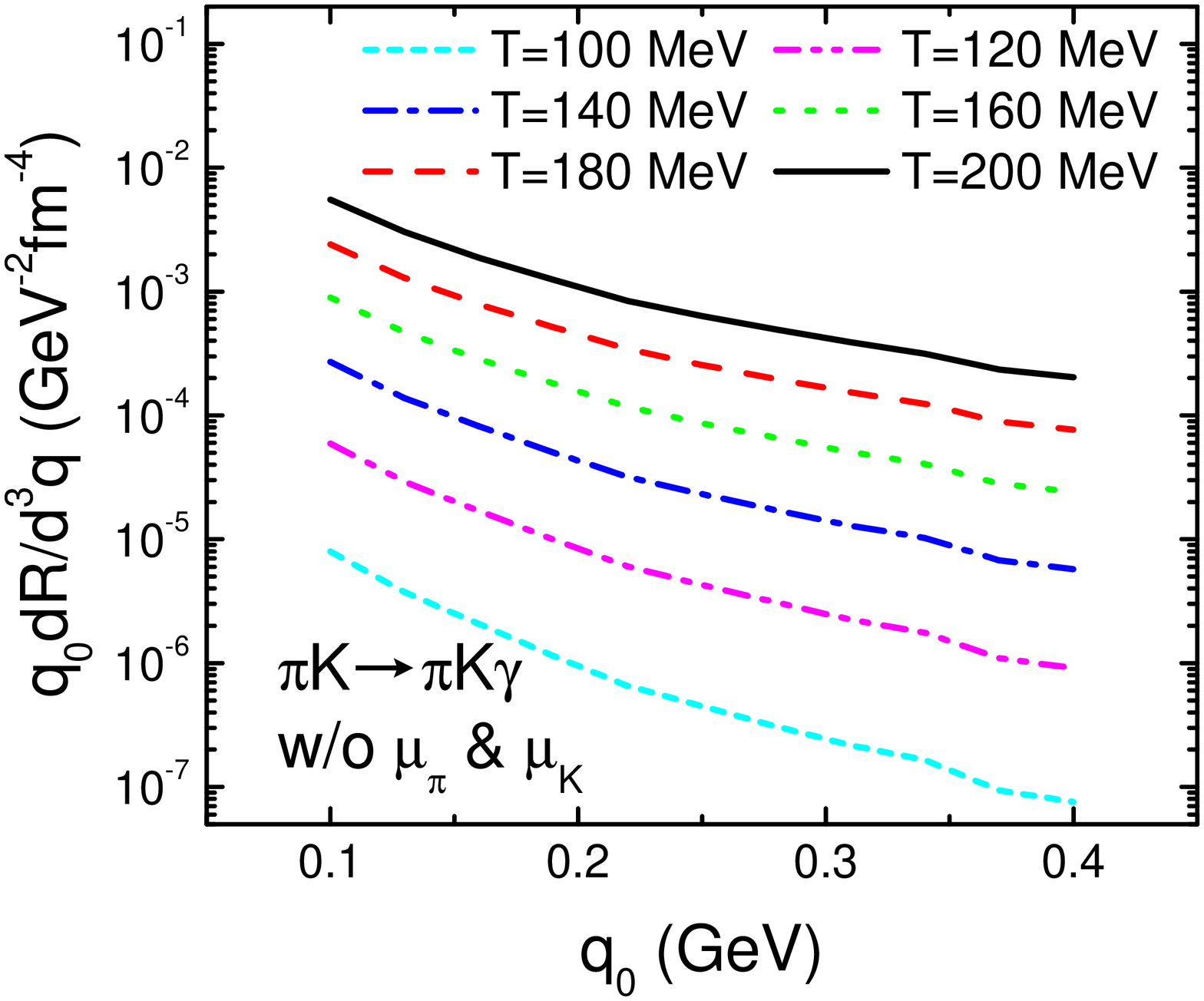}}
\caption{Thermal photon Bremsstrahlung emission rate from
$\pi K\to\pi K\gamma$ reactions as a function of photon energy at
different temperatures. The left panel includes pion- and kaon-chemical
potentials according to Eq.~(\ref{chem}), while in the right
panel $\mu_\pi=\mu_K=0$.
}
\label{fig-rateka}
\end{figure}
Turning to $\gamma$ radiation off elastic $\pi$-$K$ scattering
(Fig.~\ref{fig-rateka}), we find the emission rate to be at the
$\sim$20\% level of the one from $\pi\pi$ scattering, which essentially
reflects the experimentally observed $K/\pi$ ratio. Since $\pi$-$\pi$
scattering could provide the dominant contribution to the overall
low-energy yield (i.e., in the low-$q_t$ spectra of URHICs),
the strangeness component may be significant.
Also, the role of the kaon-chemical potential, $\mu_K$, is equally
relevant for maintaining a substantial emission rate for $\pi$-$K$
scattering toward lower temperatures (compare left and right panels
of Fig.~\ref{fig-rateka}) as is $\mu_\pi$ for $\pi$-$\pi$ scattering.

Additional sources of photon Bremsstrahlung in meson-meson scattering
have been examined in Ref.~\cite{Haglin:2003sh} within the soft photon
approximation. At a temperature of $T=200$~MeV and at low energies,
$q_0\le 0.5~GeV$, Bremsstrahlung off $\pi\pi$ scattering was found to
exceed the one off $\pi K$ scattering by about a factor of $\sim$5,
roughly in line with our results (we find a factor of $\sim$3-4).
Furthermore, $\pi\rho\to\pi\rho\gamma$ processes were found to
give a low-energy photon yield comparable to $\pi K\to \pi K\gamma$,
while other channels ($KK$, $K\rho$, $\pi K^*$) are suppressed by an
order of magnitude or more. However, at lower temperatures, the rate
for $\pi\rho\to\pi\rho\gamma$ will decrease faster than the one for
$\pi K\to\pi K\gamma$, due to thermal suppression caused by the larger
$\rho$-mass relative to the kaon mass (e.g., at T=120~MeV we estimate
the relative suppression of former relative to the latter to be a
factor of $\sim$1.7, which includes the effects of the meson-chemical
potentials; without the latter, the suppression will be larger).
$\pi\eta$ channels are suppressed relative to $\pi K$, due
to the mass and small degeneracy of the $\eta$, which, in addition,
is electrically neutral.
Thus, for meson-meson Bremsstrahlung in heavy-ion collisions, one
can expect all channels other than $\pi\pi$ and $\pi K$ to contribute on
the $\sim$20\% level or less.

Finally, let us compare the absolute magnitude of our rates to other
calculations available in the literature. For $T=200$~MeV, our
$\pi\pi\to\pi\pi\gamma$ rate for $\mu_\pi=0$ closely follows the result
of Ref.~\cite{roy96} (obtained in SPA), only that our rate is a factor
2 larger. This is precisely the enhancement we deduced from
Fig.~\ref{fig_comp} as being due to the improvements in the present work.
On the other hand, our results do not compare well with those found in
Ref.~\cite{Haglin:2003sh}, Fig.~3, where at $q_0=250$~MeV and $T=200$~MeV
the rates are larger by at least a factor of $\sim$2, despite being
evaluated in SPA. The origin of this discrepancy deserves further study.
%for $\pi\pi\to\pi\pi\gamma$ ($\pi K\to\pi K\gamma$)
%On the other, compared to the $\pi\pi\gamma$

%\begin{figure}[th]
%\includegraphics[height=4.0in,width=3.5in,angle=-90]{rate_muen.eps}
%\caption{Thermal photon bremsstrahlung radiation from $\pi\pi$ scattering
%at T=150 MeV.
%Solid line: bremsstrahlung formula (\ref{rate}) with $(1+f_1)(1+f_2)=1$;
%dashed line: $\mu_\pi=0$; dotted line: $(1+f_1)(1+f_2)=1$ and $\mu_\pi=0$;
%dash-dotted line: complete calculation.}
%\label{rate_muen}
%\end{figure}

%%%%%%%%%%%%%%%%%%%%%%%%%%%%%%%%%%%%%%%%%%%%%%%%%%%%%%%%%%%%%%%%%%%%%%%
\section{Further Uncertainties in the Emission Rates}
\label{sec_uncert}
%%%%%%%%%%%%%%%%%%%%%%%%%%%%%%%%%%%%%%%%%%%%%%%%%%%%%%%%%%%%%%%%%%%%%%
Before we apply our Bremsstrahlung rate to estimate low-energy photon
production in heavy-ion reactions at the SPS, we would
like to illustrate two additional uncertainties in the determination
of the emission rates. The first one relates to the well-known
Landau-Pomeranchuk-Migdal (LPM) interferences, while the second one
concerns possible medium effects on the meson exchanges which build
up the meson-meson interactions.

%%%%%%%%%%%%%%%%%%%%%%%%%%%%%%%%%%%%%%%%%%%%%%%%%%%%%%%%%%%
\subsection{Landau-Pomeranchuk-Migdal Effect}
\label{ssec_lpm}
\begin{figure}[!t]
\includegraphics[height=8cm,width=7.5cm,angle=-90]{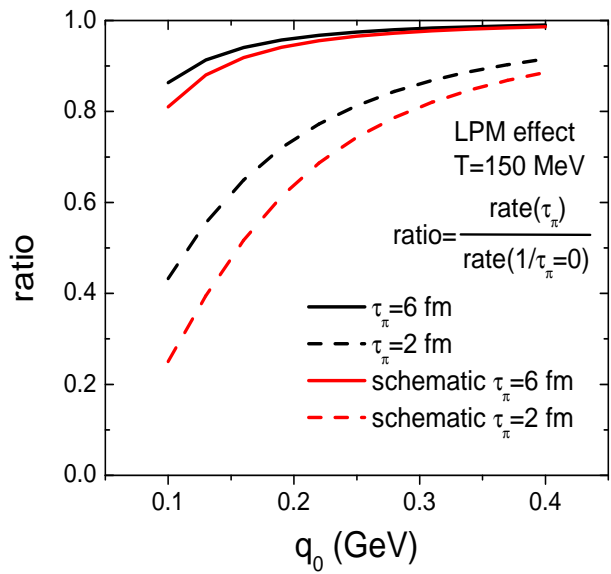}
\caption{Influence of the LPM effect on thermal photon bremsstrahlung
from $\pi\pi$ scattering at $T=150$~MeV.
}
\label{fig_lpm}
\end{figure}

%%%%%%%%%%%%%%%%%%%%%%%%%%%%%%%%%%%%%%%%%%%%%%%%%%%%%%%%%%%
When the formation time of the emitted photon, $\tau_{form}\sim 1/q_0$,
becomes of the same order as the time between subsequent $\pi$-$\pi$
collisions, $\tau_\pi^{coll}$, coherence effects are expected to
become important (LPM effect~\cite{lpm}). For soft photon
and dilepton production in a dense hadronic gas, the LPM interference
has been studied in Refs.~\cite{Cleymans93,Golov93,Knoll93}.
Here, we put these estimates into context with our updated
rate calculations.
First, we recall the expression derived in Ref.~\cite{Cleymans93},
where the LPM effect was implemented in SPA resulting in a differential
cross section of photon production off $\pi$-$\pi$ scattering as
\begin{eqnarray}
q_0\left\langle\frac{d\sigma^\gamma}{d^3q}\right\rangle
=\sigma_{\pi\pi}\frac{2\alpha}{(2\pi)^2}
\left\langle v^2\frac{(1-cos^2\theta)}{a^2+q_0^2(1-v cos\theta)^2}
\right\rangle \ .
\label{lpm}
\end{eqnarray}
$\langle \cdot \rangle$ indicates averaging over the time
between successive collisions and over all the velocities after the
collision (for simplicity, we choose an average velocity $v=0.5$c),
$\sigma_{\pi\pi}$ denotes the elastic $\pi$-$\pi$ cross section
(which factorizes in SPA) and $1/a=\tau_\pi^{coll}=1/\Gamma_\pi$
is the pion collision time (or inverse scattering width). The latter,
in turn, depends on the hadron densities and pertinent pion cross
sections, schematically given by
\begin{equation}
\Gamma_\pi = \sum\limits_h n_h(T,\mu_h)~\sigma_{\pi h}~v_{rel} \ ,
\end{equation}
where the summation is, in principle, over all hadrons in the heat bath
($h=\pi, K, N, \Delta, \dots$; $v_{rel}$: relative velocity between $\pi$
and $h$). For a selfconsistently calculated
pion gas at $(T,\mu_\pi)=(150,0)$~MeV with realistic $\pi\pi$
interactions~\cite{Rapp:1993ih}, one finds $\tau_\pi^{coll}\simeq 6$~fm/c,
while a hot $\pi N\Delta$ gas representative for SPS conditions
leads to $\tau_\pi^{coll}\simeq 2$~fm/c~\cite{Rapp:1993bi}.
In Fig.~\ref{fig_lpm} we used both values to bracket the uncertainty
in the suppression of low-energy photons, by plotting the ratio
of Eq.~(\ref{lpm}) to the incoherent limit ($a=0$).
For $\tau_\pi^{coll}=6$~fm/c, the quenching factor is rather small
reducing the photon emission rate by no more than 15\% even at the
smallest energies considered here, $q_0=100$~MeV. This is consistent
with Ref.~\cite{Knoll93} where a slightly different coherence
factor has been derived, amounting to a multiplication of the incoherent
rate by  $[(q_0\tau)^2/(1+(q_0\tau)^2)]^2$; for $\tau_\pi^{coll}=6$~fm/c
and $q_0=100$~MeV the latter turns out to be 0.81.
However, for $\tau_\pi^{coll}=2$~fm/c and $q_0=100$~MeV, collision and
formation time are of comparable magnitude, and the quenching factor
becomes substantial, between 0.4 and 0.25 according to the two
ways of estimating it.

%%%%%%%%%%%%%%%%%%%%%%%%%%%%%%%%%%%%%%%%%%%%%%%%%%%%%%%%%%%
\subsection{Schematic Estimate of Medium Effects}
\label{ssec_med}
%%%%%%%%%%%%%%%%%%%%%%%%%%%%%%%%%%%%%%%%%%%%%%%%%%%%%%%%%%%
When hadrons are embedded into a hot and dense medium, their vacuum
properties (mass and width) are expected to change, which is
encoded in pertinent spectral functions,
\begin{equation}
A_h(k_0,k)= -2~{\rm Im} D_h(k^2)=-2~{\rm Im} \left(
\frac{1}{k^2-m_h^2-\Sigma_h(k_0,k)} \right) \ ,
\end{equation}
where $\Sigma_h(k_0,k)$ is the self-energy of meson $h$ as induced
by interactions with the surrounding matter particles.
For the present purpose, the relevant mesons are the $\rho$ and
the ``$\sigma$" which figure into the meson-exchange propagators
in the amplitudes (since we work in Born approximation, no explicit
pion propagators appear).
Dilepton spectra in heavy-ion collisions have revealed ample evidence
that the in-medium $\rho$-meson spectral function is substantially
modified. While the low-mass enhancement observed by CERES/NA45 in
$Pb$(158~AGeV)-$Au$ collisions~\cite{Agakichiev:2005ai} is compatible with
both the assumption of a dropping $\rho$-mass and a broadening as
predicted by hadronic many-body theory~\cite{RW99}, more recent NA60
dimuon spectra~\cite{Arnaldi:2006jq} clearly favor the
latter~\cite{vanHees:2006ng}. Concerning the ``$\sigma$" meson,
2-pion production experiments off
nuclei~\cite{Grion:2005hu,Starostin:2000cb} indicate substantial medium
effects as well, in terms of an enhancement
near the two-pion threshold which increases with nuclear mass number $A$.
Whether this is due to a dropping ``$\sigma$"-meson mass~\cite{Hats99}
or hadronic many-body effects~\cite{Cabrera:2005wz} (or both), is not
clear at present. At finite temperature, the (the approach to)
chiral symmetry restoration (implying degeneracy of $\pi$- and
$\sigma$- spectral functions), is suggestive for a $\sigma$-mass
approaching the pion mass~\cite{Roder:2005qy}.
It is therefore desirable to investigate to what
extent Bremsstrahlung off $\pi$-$\pi$ scattering is sensitive to
modifications of ``$\sigma$" and $\rho$ mesons. Previous studies along
these lines~\cite{Halasz:1997xc,Song:1998um,Alam:2001ar} have mostly
addressed hadronic photon emission at higher energies (in connection
with a dropping $\rho$ mass).
%which could be tested against available data from WA80 and WA98.

In light of the above, we here focus on the effects of a reduced
$\sigma$-mass on low-energy photon emission. We will not address
changes in the $\rho$-meson properties, as a simple dropping mass
appears to be ruled out while a consistent implementation of changes
in the width in the propagator requires accounting for the pertinent
processes in the photon production rate as well (for the same
reason, we do not consider width changes in the $\sigma$-propagator).

\begin{figure}[t]
\centerline{
%\includegraphics[height=8.0cm,width=8.0cm,angle=-90]{medium1.eps}
%\hspace{-1.0cm}
\includegraphics[height=9.5cm,width=8.5cm,angle=-90]{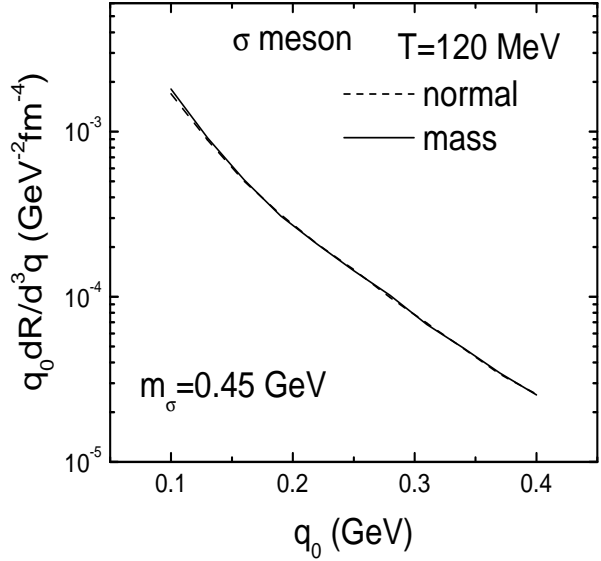}}
\caption{In-medium effects on thermal photon production implemented by
a dropping ``$\sigma$"-meson mass in Bremsstrahlung off $\pi$-$\pi$
scattering.  Dashed ($m_\sigma=0.525$~GeV) and solid line
($m_\sigma=0.45$~GeV) are almost indistinguishable due to the
minor role of ``$\sigma$"-exchange in the total elastic
$\pi$-$\pi$ cross section at low energy.}
\label{fig_med}
\end{figure}
When recalculating photon Bremsstrahlung off $\pi$-$\pi$ scattering
by reducing the mass of the $\sigma$ meson to $m_\sigma=0.45$~GeV
(as compared to 0.525~GeV before), the thermal emission rate
barely changes, cf.~Fig.~\ref{fig_med}.
This can be readily understood by recalling in the elastic $\pi$-$\pi$
cross section the $\sigma$-meson contribution plays a minor role,
cf.~right panel in Fig.~\ref{fig_sigpp}.
Nevertheless, the smallness of the effect when dropping the $\sigma$ mass
by $\sim$15\% is not necessarily expected, but forces us to conclude
that low-energy thermal photon spectra are not sensitive to
(partial) chiral symmetry restoration in the scalar channel
(for completeness we state that the rate of low-energy photon production
increases by $\sim$ 25\% if one were to reduce the $\rho$-meson mass
from its vacuum value of 0.77~GeV to 0.65~GeV).

%%%%%%%%%%%%%%%%%%%%%%%%%%%%%%%%%%%%%%%%%%%%%%%%%%%%%%%%
\section{Photon Spectra in Heavy-Ion Collisions}
\label{sec_spec}
%%%%%%%%%%%%%%%%%%%%%%%%%%%%%%%%%%%%%%%%%%%%%%%%%%%%%%%%
We now apply our Bremsstrahlung rates to the calculation of
photon spectra at low transverse momentum in central
$Pb$(158~AGeV)-$Pb$ collisions at SPS. We supplement the spectra
with the baseline results of Ref.~\cite{TRG04}\footnote{To avoid double
counting, we have removed the contributions from $\rho\to\pi\pi\gamma$,
$\pi\pi\to\rho\gamma$, $K^*\to\ K\pi\gamma$ and $\pi K\to K^*\gamma$
from the rates of Ref.~\cite{TRG04}, as they are part of the
$\pi\pi\to\pi\pi\gamma$ and $\pi\pi\to\pi\pi\gamma$ processes
when using full $\pi\pi$ and $\pi K$ cross sections;
the latter include (unstable) $\rho$ and $K^*$ mesons which should
not be counted twice. \label{fn_dc}}
and compare the total to the recent WA98 data~\cite{wa98-04}.
As in Ref.~\cite{TRG04}, the folding of the
thermal emission rates over the space-time evolution is performed
with the help of an expanding fireball evolution which has been
parametrized to reflect basic features of hydrodynamic models.
With a hadronic resonance-gas equation of state (with chemical
freezeout at $(\mu_N^{ch},T_{ch})=(250,175)$~MeV) and
a participant-nucleon number of $N_{part}=340$ to represent
the 10\% most central collisions, the fireball volume at chemical
freezeout is fixed
to approximately reproduce the observed hadron multiplicities (after
chemical freezeout, the evolution is constructed using entropy and
baryon-number conservation, as well as meson-chemical potentials
to preserve the correct hadron ratios).
After averaging over the appropriate rapidity interval of the
WA98 acceptance the thermal photon spectra follow as
\begin{eqnarray}
q_0\frac{dN}{d^3q}(q_t)=\frac{1}{\Delta y}
\int\limits_{y_{min}}^{y_{max}}dy\int
d\tau V_{FB}(\tau)q_0\frac{dR^\gamma}{d^3q}
\end{eqnarray}
with $y_{min}=2.35$ and $y_{max}=2.95$ ($\Delta y=0.6$). The fireball
volume is taken to evolve with proper time as
\begin{equation}
V_{FB}(\tau)=\frac{1}{2}\pi(z_0+v_{z,0}\tau+\frac{1}{2}a_z\tau^2)
(r_{T,0}+\frac{1}{2}a_T\tau^2)^2
\end{equation}
with $z_0=1.8$~fm (corresponding to a formation time of 1~fm/c and
translating into an average initial temperature of $T_0=205$~MeV),
$r_{T,0}=6.25$~fm, $a_z=a_T=0.045$~c$^2$/fm, and $v_z=0.6$c (note that
only half of the total fireball volume must be used to reproduce
the correct $dN_{ch}/dy$).
\begin{figure}[!tb]
\centerline{
\includegraphics[height=8.0cm,width=8.0cm,angle=0]{wa98-low-23.eps}
\hspace{1.0cm}
\includegraphics[height=7.8cm,width=8.0cm,angle=0]{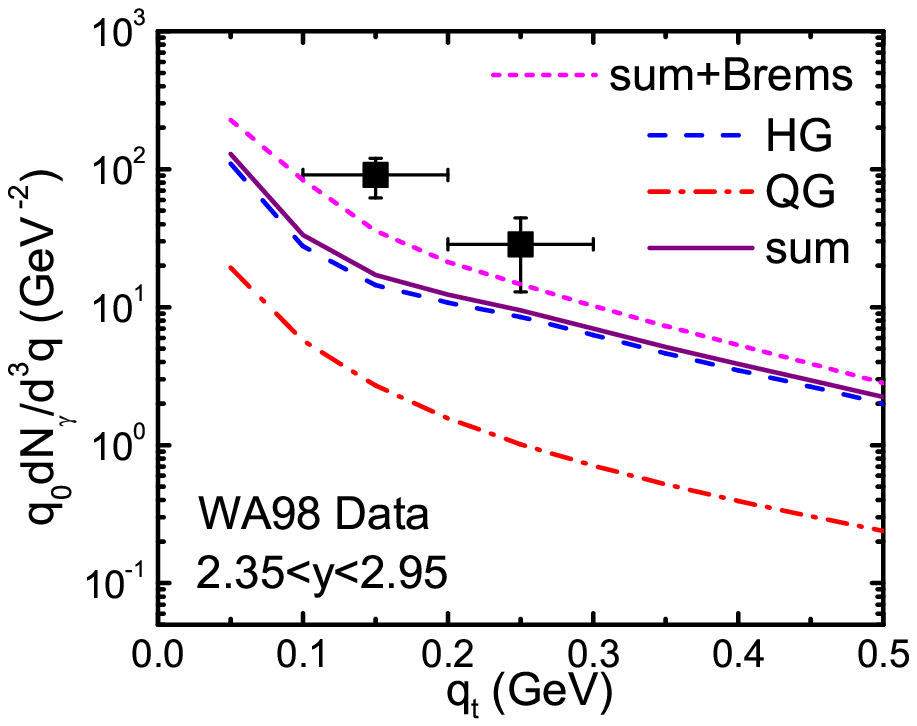}}
\caption{Direct low-$q_t$ photon spectra as measured in central
$Pb$-$Pb$ collisions at SPS~\cite{wa98-04} compared to thermal emission
spectra from an expanding thermal fireball with QGP and hadronic
phases~\cite{TRG04}. In the left panel (taken from
Refs.~\cite{Tur04,Rapp:2005fy}) QGP~\cite{AMY01} (dash-dotted line) and
hadron gas (HG) emission~\cite{TRG04} (dashed line, without meson-meson
Bremsstrahlung) add up to the solid line, while the upper dashed line
additionally includes an estimate~\cite{Tur04} of Bremsstrahlung off
$\pi$-$\pi$ scattering in the soft-photon approximation (SPA) and
without final-state Bose enhancement factors, representing the rate
depicted by the solid line in Fig.~\ref{fig_comp} (the contributions
from $\rho\to\pi\pi\gamma$ decays in the HG rates of Ref.~\cite{TRG04}
have been removed to avoid double counting). The downward arrows
represent 90\% upper confidence limits set by WA98 in an earlier
measurement~\cite{wa98-00}.
The curves in the right panel are as in the left panel except that
the upper dashed line now includes the improved rates
from elastic $\pi$-$\pi$ and $\pi$-$K$ Bremsstrahlung (without SPA,
including final-state Bose enhancement and with a similar
caveat on double counting as in the left panel,
cf.~footnote~\ref{fn_dc}).}
\label{fig_spec}
\end{figure}
QGP and mixed phase are completed after $\tau_H\simeq 4.46$~fm/c and
thermal freezeout is assumed at $\tau_{fo}=13$~fm/c (plus one extra
fm/c to allow for strong resonance decays). The numerically inferred
temperature evolution in the hadronic phase can be conveniently
parametrized as
\begin{equation}
T(\tau)=\frac{1}{c_1+c_2R(\tau)} {\rm GeV} \ ,
\end{equation}
with $c_1$=1.3, $c_2$=0.639/fm, and a ``radius"
\begin{equation}
R(\tau)=\left(\frac{3}{4\pi}V_{FB}(\tau)\right)^{\frac{1}{3}} \ .
\end{equation}
Parameterizations of the numerically computed Bremsstrahlung rates
along the temperature-chemical potential trajectory used here
(as well as separately in $q_0$, $T$ and $\mu_\pi$)
are quoted in Appendix~\ref{app_param}.

In Fig.~\ref{fig_spec} we summarize our results for direct photon
spectra in central $Pb$(158~AGeV)-$Pb$ in comparison to the low-$q_t$
data of WA98~\cite{wa98-04} at the CERN-SPS. As noted
in the introduction, these data are extracted via a photon HBT method
requiring an assumption on the emission source size implying an extra
(systematic) error of the data (in addition, strictly speaking, the
weighted average of each data point does not correspond to the center
of the transverse-momentum bin, but is shifted to somewhat
smaller $q_t$, due to the rather sharp increase of the yield toward
lower $q_t$).
The left panel of Fig.~\ref{fig_spec} reproduces the previous
calculations for QGP and hadron
gas emission of Ref.~\cite{TRG04} (lower three curves),
which are in good agreement with earlier WA98 spectra at
$q_t\ge1.5$~GeV, but significantly underestimate the low-momentum
data at $q_t=0.1-0.3$~GeV. This is still true upon inclusion of
Bremsstrahlung off $\pi$-$\pi$ scattering~\cite{Tur04,Rapp:2005fy}
treated in SPA and without final-state Bose enhancement, even
though this contribution is noticeable.
However, if the improved Bremsstrahlung rates from $\pi$-$\pi$ and
$\pi$-$K$ interactions are employed, the discrepancy is reduced
appreciably, see right panel of Fig.~\ref{fig_spec} (note that the
Bremsstrahlung yield decreases faster with energy than the other
hadronic contributions, thereby not upsetting the agreement of the
previous calculations with the experimental spectra (upper limits)
at higher energies).
These findings are presumably to be considered as an upper limit
since at this point we do not account for LPM suppression effects
as discussed in Sec.~\ref{ssec_lpm}.
The thermal low-$q_t$ photon yield is furthermore found to
approximately scale with the fireball lifetime, which reiterates
the importance of the later hadronic stages (on the other hand,
the sensitivity to transverse flow effects is small).

We also note that
baryon-induced effects, which enter through a medium
modified $\rho$ spectral function~\cite{RW99} carried to the photon
point and which dominate the contribution labeled ``HG" at the
depicted momenta,
are appreciable (at the 30-60\% level in the range $q_t=0.1-0.3$~GeV).
Besides resonance decays
(e.g., $\Delta, N(1520)\to N\gamma$), these include Bremsstrahlung
from $\pi$-$N$, $N$-$N$, and $N$-$\Delta$ interactions (as following
from $\pi NN^{-1}$ and $\pi\Delta N^{-1}$ excitations in the pion
cloud of the $\rho$-meson), which have been constrained by
(the inverse process of) photoabsorption on the nucleon and
nuclei~\cite{Rapp:1997ei}. In Ref.~\cite{alam03}, similar
contributions were claimed to be smaller.

%%%%%%%%%%%%%%%%%%%%%%%%%%%%%%%%%%%%%%%%%%%%%%%%%%%%%%%%%%%%%%%%%%%
\section{Coherent Emission from Heavy-Ion Collisions}
\label{sec_coh}
%%%%%%%%%%%%%%%%%%%%%%%%%%%%%%%%%%%%%%%%%%%%%%%%%%%%%%%%%%%%%%%%%%%
\begin{figure}[tb]
\vspace{0.5cm}
\centerline{
\includegraphics[height=7.0cm,width=7.0cm,angle=-90]{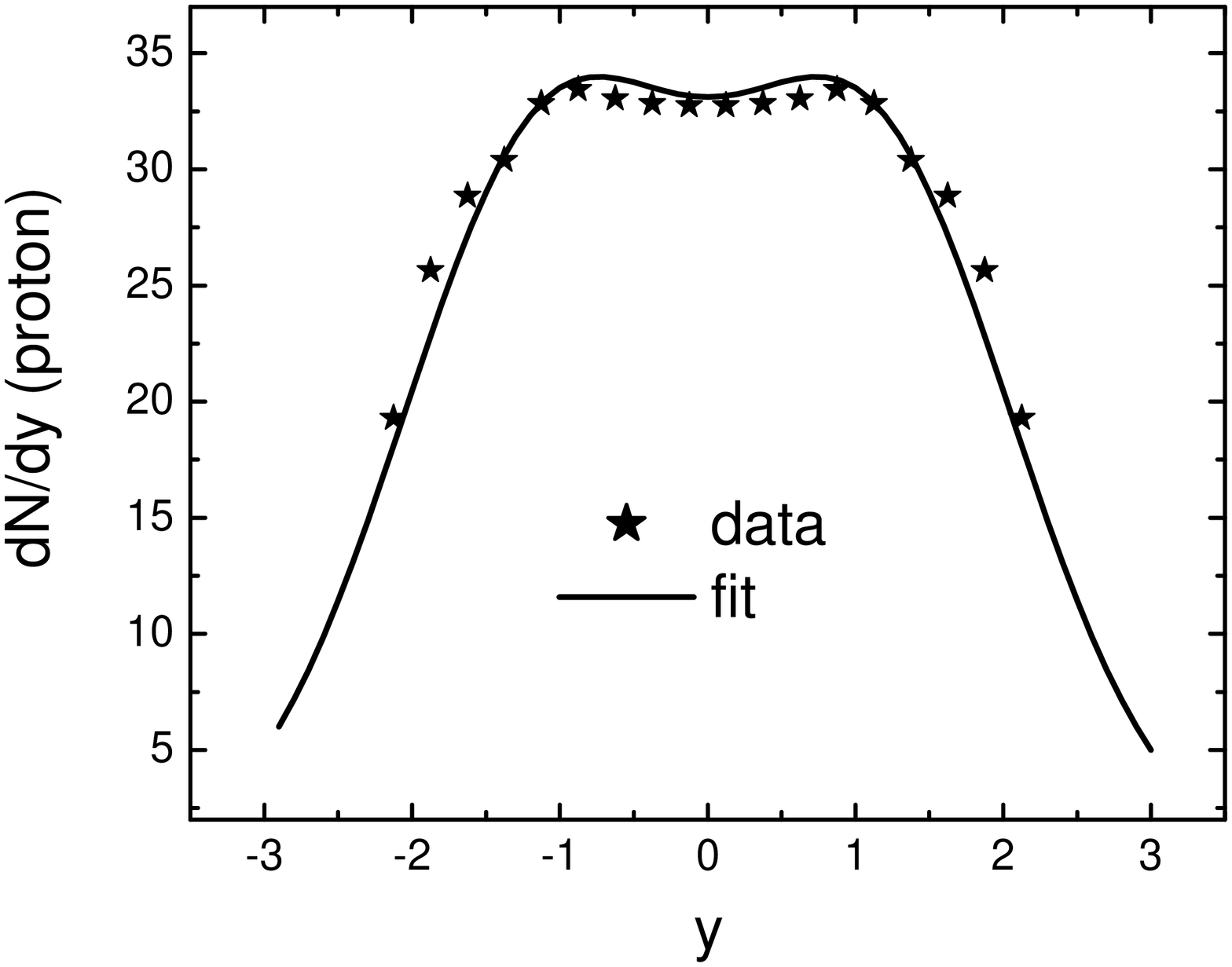}
\hspace{0.8cm}
\includegraphics[height=7.0cm,width=7.0cm,angle=-90]{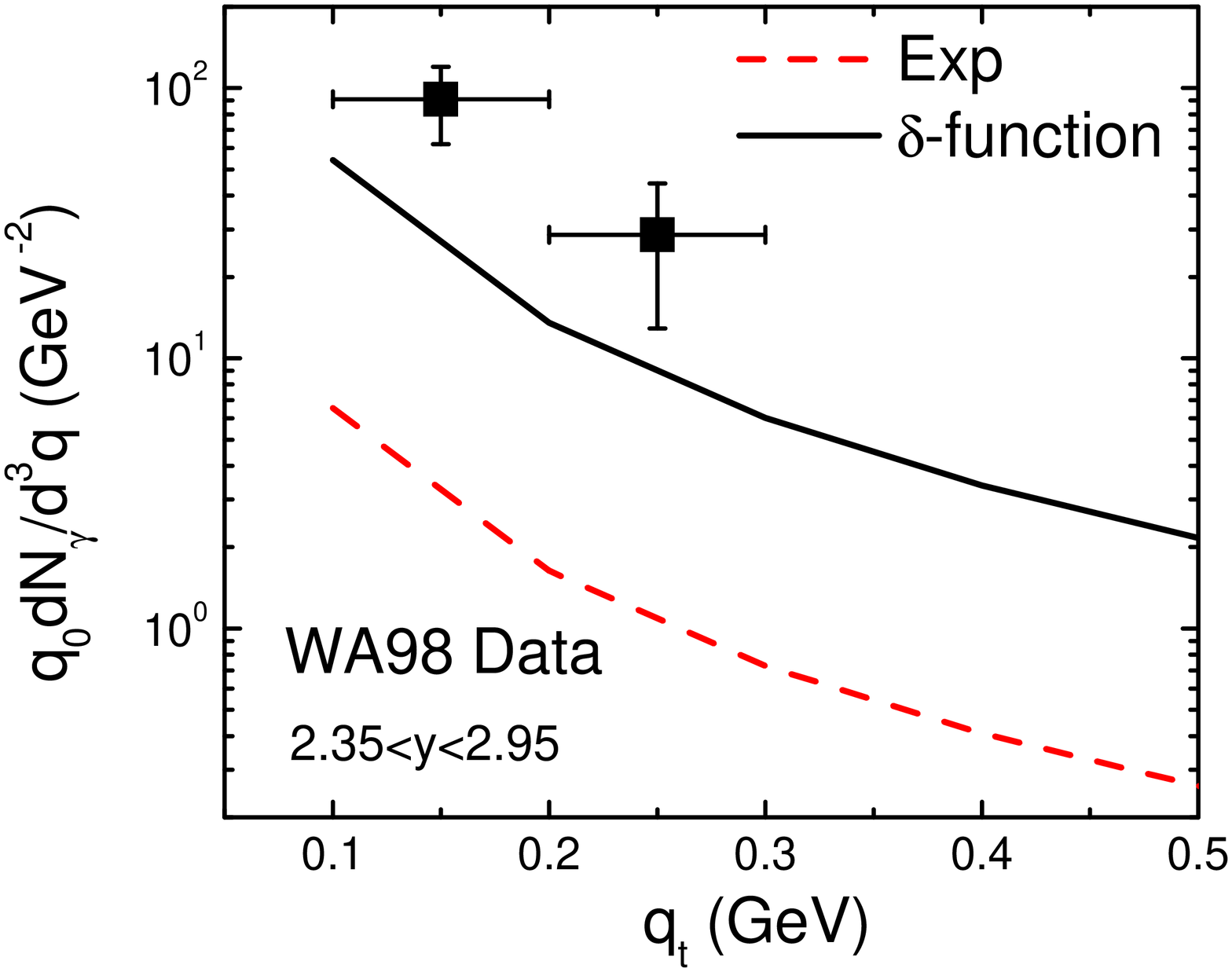}}
\caption{Left panel: two-Gaussian fit (with centers at $y_0=\pm 1.1$ and
width $\sigma_y=1.0$) to the experimental proton rapidity
distributions~\cite{na49} in central $Pb$(158~AGeV)-$Pb$.
Right panel: coherent photon spectra from $Pb$-$Pb$ collisions at SPS
using final-state proton rapidity distributions from experiment
(dashed line) or $\delta$-functions centered at $y_{P,T}=\pm 0$
(solid line), compared to the experimental data~\cite{wa98-04}.
}
\label{fig_coh}
\end{figure}
Another source of soft direct photons in heavy-ion collisions that we
would like to study here is (nonthermal) Bremsstrahlung emission due
to coherent radiation off the protons within the incoming
nuclei~\cite{bjorken,Ko86}. In the long-wave length limit,
$\lambda_\gamma=2\pi/q \gg D$ ($q$: photon 3-momentum, $D\simeq10$~fm:
typical system size), phase interferences are suppressed and the
radiation can be assessed via the rapidity shift that the protons
undergo from the initial nuclei to the finally observed distribution.
Including both target and projectile, and for central collisions, one
finds the following form of the photon spectrum~\cite{bjorken,Ko86},
\begin{eqnarray}
q_0\frac{dN_\gamma}{d^3q}=\frac{\alpha_{em}}{4\pi^2 q^2_0}
\left|\int dy\left\{\left.\frac{dN}{dy}\right|_P
\left[\frac{v{\rm sin\theta}}{1-v{\rm cos\theta}}
-\frac{v_P {\rm sin\theta}}{1-v_P {\rm cos\theta}}\right]
+\left.\frac{dN}{dy}\right|_T
\frac{v {\rm sin\theta}}{1-v {\rm cos\theta}}\right\}\right|^2
\label{co_equation}
\end{eqnarray}
where $v_P$ is the beam velocity, $y=\tanh^{-1}v$ is the
final-state (FS) proton rapidity, and
\begin{eqnarray}
\left.\frac{dN}{dy}\right|_P ~~{\rm and} ~~ \left.\frac{dN}{dy}\right|_T
\end{eqnarray}
are the pertinent FS rapidity distributions for projectile and target
nuclei, respectively. The latter can be reasonably well fitted to
experimental data~\cite{na49} using two Gaussians, cf.~left
panel of Fig.~\ref{fig_coh}.
In the right panel of Fig.~\ref{fig_coh} we summarize our results
for coherent emission from the longitudinal deceleration of the incoming
protons. With realistic final rapidity distributions (dashed line) the
photon yield constitutes only a few percent of the WA98 low-energy data,
which presumably is an upper limit since already for $q_0=0.1$~GeV, the
pertinent photon wavelength, $\lambda_\gamma\simeq12$~fm, is of the
order of the system size rendering phase variations (interference)
potentially relevant. Even with the extreme assumption of
$\delta$-function like proton rapidity distributions at $y_{P,T}=0$
(full stopping), the coherent emission yield falls significantly short
of the data. We thus conclude that this mechanism cannot
be at the origin of the observed enhancement over previous
predictions~\cite{TRG04}.

%%%%%%%%%%%%%%%%%%%%%%%%%%%%%%%%%%%%%%%%%%%%%%%%%%%%%
\section{Summary and Discussion}
\label{sec_sum}
%%%%%%%%%%%%%%%%%%%%%%%%%%%%%%%%%%%%%%%%%%%%%%%%%%%%%
In this article we have revisited the problem of soft
photon production from hadronic matter which has
recently received renewed interest by the observation of a large
excess by the WA98 collaboration in central $Pb$-$Pb$ collisions
at the CERN-SPS.
Earlier predictions using two-body mesonic and baryon-induced thermal
photon production significantly underestimated these data.
We have focused on a more elaborate re-evaluation of
Bremsstrahlung from meson-meson interactions. Based on a
${\rm U_{\rm em}}(1)$-gauged meson-exchange model for elastic
$\pi$-$\pi$ and $\pi$-$K$ scattering (in Born approximation), our main
improvement over previous calculations consists of a full numerical
treatment of the collision integral which goes beyond the commonly
employed soft-photon approximation and allows for the inclusion
of final-state Bose enhancement factors.
We have explicitly constructed pertinent contact interactions to
preserve electromagnetic gauge invariance of the underlying
Bremsstrahlung amplitude.
The combined effect of the two improvements amounts to an increase
of the thermal $\pi$-$\pi$ Bremsstrahlung rate by up to a factor
of 2 at low photon energies ($q_0=0.1-0.5$~GeV), while $\pi$-$K$
scattering adds another $\sim$20\% (essentially reflecting the
$K$/$\pi$ ratio in the hadronic gas). As a consequence, meson-meson
Bremsstrahlung becomes the dominant thermal photon source at low
energies. When combined with previous rate calculations, the
convolution over a standard thermal fireball evolution
reduces previously found discrepancies with the low-momentum WA98
data. This comparison,
however, does not include the LPM effect, which we have estimated to
suppress the soft emission rate by up to several tens of percent.
We have furthermore shown that the Bremsstrahlung rates are rather
insensitive to medium effects on the "$\sigma$"-meson in form
of a reduced mass, due to the fact that the elastic $\pi$-$\pi$
cross section is dominated by ($s$-channel) $\rho$ exchange.
We finally studied the possibility of coherent Bremsstrahlung
induced by rapidity shifts of the incoming projectile and
target protons. When using realistic final-state distributions,
we have found this contribution to be essentially negligible.

While our investigations suggest that thermal emission from hadronic
matter might be at the origin of the low-$q_t$ WA98 enhancement,
more work is required to quantify the theoretical estimates,
e.g.~by merging medium effects on the $\rho$-meson (as implicit
in rate estimates using in-medium spectral functions)
with the newly computed Bremsstrahlung
rates, as well as by explicit implementations of the LPM effect.
In addition, an experimental confirmation of the observed enhancement,
e.g.~at the Relativistic Heavy-Ion Collider (RHIC), would be very
illuminating.

\begin{acknowledgments}
We gratefully acknowledge discussions with C.~Gale and S.~Turbide in the
early stages of this work, and for supplying us with the Bremsstrahlung
rates used in the left panel of Fig.~\ref{fig_spec}.
We also wish to thank Che-Ming Ko and Hendrik van Hees
for helpful discussions.
This work was supported in part by a U.S. National Science Foundation
CAREER award under grant PHY-0449489.
\end{acknowledgments}

%%%%%%%%%%%%%%%%%%%%%%%%%%%%%%%%%%%%%%%%%%%%%%%%%%%%%%%%%%%%%%%%%%%%%%
\appendix

%%%%%%%%%%%%%%%%%%%%%%%%%%%%%%%%%%%%
\section{Evaluation of the Bremsstrahlung Phase Space Integral}
\label{app_integral}
%%%%%%%%%%%%%%%%%%%%%%%%%%%%%%%%%%%%
In this appendix, we derive the expressions for the photon emission
rate as used in Sec.~\ref{sec_rate}.
Starting from the kinetic theory expression, Eq.~(\ref{rate0}) in the
main text, for $\pi+\pi\to\pi+\pi+\gamma$,
\begin{eqnarray}
q_0\frac{dR^\gamma}{d^3q}&=&N\int\frac{d^3p_a}{2E_a(2\pi)^3}
\frac{d^3p_b}{2E_b(2\pi)^3}
\frac{d^3p_1}{2E_1(2\pi)^3}\frac{d^3p_2}{2E_2(2\pi)^3}\nonumber\\
&&\times(2\pi)^4\delta^4(p_a+p_b-p_1-p_2-q)|{\cal M}_i|^2
\frac{f_a(E_a)f_b(E_b)[1+f_1(E_1)][1+f_2(E_2)]}{2(2\pi)^3} \ ,
\end{eqnarray}
we perform the integrations over the energy-momentum conserving delta
function to obtain
\begin{eqnarray}
q_0\frac{dR^\gamma}{d^3q}&=&\frac{N}{16(2\pi)^{11}}\int
\frac{d^3p_a}{E_a}\frac{d^3p_b}{E_b}
\int \frac{d^3p_1}{E_1}f_1f_2[1+f_3][1+f_4]\nonumber\\
&&\times |{\cal M}_i|^2\delta((E_a+E_b-E_1-q_0)^2
-({\bf p}_a+{\bf p}_b-{\bf p}_1-{\bf q})^2-m^2_4)\nonumber\\
&=&\frac{N}{16(2\pi)^{11}}\int |{\bf p}_a|dE_ad\Omega_a\int|{\bf p}_b|dE_bd\Omega_b\int|
{\bf p}_1|dE_1d\Omega_1f_1f_2[1+f_3][1+f_4]\nonumber\\
&&\times |{\cal M}_i|^2\delta((E_a+E_b-E_1-q_0)^2
-({\bf p}_a+{\bf p}_b-{\bf p}_1-{\bf q})^2-m^2_2) \ .
\label{rate2}
\end{eqnarray}
We evaluate this equation by the Monte Carlo method, choosing
the coordinates for each particle as follows,
\begin{eqnarray}
p_a&=&(E_a, |{\bf p}_a|{\rm sin\theta_acos\phi_a,
|{\bf p}_a|sin\theta_asin\phi_a,|{\bf p}_a|cos\theta_a}),\nonumber\\
p_b&=&(E_b, |{\bf p}_b|{\rm sin\theta_bcos\phi_b,
|{\bf p}_b|sin\theta_bsin\phi_b,|{\bf p}_b|cos\theta_b}),\nonumber\\
p_1&=&(E_1, |{\bf p}_1|{\rm sin\theta_1cos\phi_1,
|{\bf p}_1|sin\theta_1sin\phi_1,|{\bf p}_1|cos\theta_1}),\nonumber\\
q&=&(q_0, 0,0,|{\bf q}|),\nonumber\\
p_2&=&p_a+p_b-p_1-q \ ,
\end{eqnarray}
with
\begin{eqnarray}
E_i&=&\sqrt{m_i^2+{\bf p}_i^2} \quad (i=a,b,1,2,q) \ .
\end{eqnarray}
Thus, Eq.~(\ref{rate2}) can be rewritten as
\begin{eqnarray}
q_0\frac{dR^\gamma}{d^3q}&=&\frac{N}{16(2\pi)^{11}}
\int |{\bf p}_a|dE_a{\rm sin\theta_a}
d{\rm \theta_a}d\phi_a\int|{\bf p}_b|dE_b{\rm sin\theta_b}
d{\rm \theta_b}d{\rm \phi_b}\int|
{\bf p}_1|dE_1{\rm sin\theta_1}d{\rm \theta_1}
d{\rm \phi_1}f_1f_2[1+f_3][1+f_4]
\nonumber\\
&&\times |{\cal M}_i|^2
\delta((E_a+E_b-E_1-q_0)^2-({\bf p}_a+{\bf p}_b-{\bf p}_1-{\bf q})^2-m^2_2).
\end{eqnarray}
In the argument of the delta function, one has
\begin{eqnarray}
({\bf p}_a+{\bf p}_b-{\bf p}_1-{\bf q})^2
&=&|{\bf p}_a|^2+2{\bf p}_a\cdot{\bf p}_b
-2{\bf p}_a\cdot{\bf p}_1-2{\bf p}_a\cdot{\bf q}+({\bf p}_b
-{\bf p}_1-{\bf q})^2
\nonumber\\
&=&2|{\bf p}_a||{\bf p}_b|{\rm sin\theta_a sin\theta_b cos\phi_acos\phi_b}
+2|{\bf p}_a||{\bf p}_b|{\rm sin\theta_a sin\theta_b sin\phi_asin\phi_b}
+2|{\bf p}_a||{\bf p}_b|{\rm cos\theta_a cos\theta_b}\nonumber\\
&&-2|{\bf p}_a||{\bf p}_1|{\rm sin\theta_a sin\theta_1 cos\phi_acos\phi_1}
-2|{\bf p}_a||{\bf p}_1|{\rm sin\theta_a sin\theta_1 sin\phi_asin\phi_1}
-2|{\bf p}_a||{\bf p}_1|{\rm cos\theta_a cos\theta_1}\nonumber\\
&&-2|{\bf p}_a||{\bf q}|{\rm cos\theta_a}
+|{\bf p}_a|^2+({\bf p}_b-{\bf p}_1-{\bf q})^2 \ .
\end{eqnarray}
Upon integrating over $\theta_a$, we arrive at
\begin{eqnarray}
q_0\frac{dR^\gamma}{d^3q}&=&\frac{N}{16(2\pi)^{11}}
\int |{\bf p}_a|dE_ad\phi_a\int|{\bf p}_b|dE_b{\rm sin\theta_b}
d{\rm \theta_b}d{\rm \phi_b}\int|
{\bf p}_1|dE_1{\rm sin\theta_1}d{\rm \theta_1}
d{\rm \phi_1}f_1f_2[1+f_3][1+f_4]\frac{|{\cal M}_i|^2}{\cal A}
\label{finalrate}
\end{eqnarray}
with
\begin{eqnarray}
{\cal A}=|\varphi'({\rm cos}\theta^r_a)| \ .
\end{eqnarray}
In the above, ${\rm cos}\theta^r_a$ is the root of function
$\varphi({\rm cos}\theta_a)$,
\begin{eqnarray}
\varphi({\rm cos}\theta_a)
&=&(2|{\bf p}_a||{\bf p}_b|{\rm sin\theta_bcos\phi_acos\phi_b}
+2|{\bf p}_a||{\bf p}_b|{\rm sin\theta_bsin\phi_asin\phi_b}
-2|{\bf p}_a||{\bf p}_1|{\rm sin\theta_1cos\phi_acos\phi_1}
\nonumber\\
&&-2|{\bf p}_a||{\bf p}_1|{\rm sin\theta_1sin\phi_asin\phi_1})
\sqrt{1-{\rm cos}^2\theta_a}
\nonumber\\
&&+(2|{\bf p}_a||{\bf p}_b|{\rm cos\theta_b}
-2|{\bf p}_a||{\bf p}_1|{\rm cos\theta_1}
-2|{\bf p}_a||{\bf q}|){\rm cos\theta_a}
\nonumber\\
&&+|{\bf p}_a|^2+({\bf p}_b-{\bf p}_1-{\bf q})^2
-(E_a+E_b-E_1-q_0)^2+m^2_2=0.\label{equation}
\end{eqnarray}
with
\begin{eqnarray}
({\bf p}_b-{\bf p}_1-{\bf q})^2&=&|{\bf p}_b|^2+|{\bf p}_1|^2+|{\bf p}_q|^2
-2|{\bf p}_b||{\bf p}_1|{\rm sin\theta_b sin\theta_1 cos\phi_bcos\phi_1}
-2|{\bf p}_b||{\bf p}_1|{\rm sin\theta_b sin\theta_1 sin\phi_bsin\phi_1}
\nonumber\\
&&-2|{\bf p}_b||{\bf p}_1|{\rm cos\theta_b cos\theta_1}
-2|{\bf p}_b||{\bf q}|{\rm cos\theta_b}+|{\bf p}_1||{\bf q}|{\rm cos\theta_1}.
\end{eqnarray}
There are two additional constraints on the eight integration variables in
Eq.~(\ref{finalrate}):
\begin{description}
\item{a.} For any configuration of the eight variables,
Eq.~(\ref{equation}) must have
at least one root.
\item{b.} $E_a+E_b\ge E_1+q_0+m_2$.
\end{description}

%%%%%%%%%%%%%%%%%%%%%%%%%%%%%%%%%%%%%%%%%%%%%%%%%%%%%%%%%%%
\section{Gauge Invariance of Photon Production Amplitudes}
\label{app_gauge}
%%%%%%%%%%%%%%%%%%%%%%%%%%%%%%%%%%%%%%%%%%%%%%%%%%%%%%%%%%%
In this appendix, we demonstrate gauge invariance for the process
$\pi^+\pi^-\to\pi^+\pi^-\gamma$. In this case the charges are,
$Q_a=Q_1=1$, $Q_b=Q_2=-1$, and there is no $u$-channel contribution.
To verify ${\cal M}^\mu\cdot q_\mu=0$ for the
$t$-channel amplitude, we have
\begin{eqnarray}
{\cal M}^\mu_t\cdot q_\mu&=&e(J^\mu_a{\cal M}_t(p_a-q,p_b,p_1,p_2)
+J^\mu_b{\cal M}_t(p_a,p_b-q,p_1,p_2)
\nonumber\\
&&+J^\mu_1{\cal M}_t(p_a,p_b,p_1+q,p_2)
+J^\mu_2{\cal M}_t(p_a,p_b,p_1,p_2+q))\cdot q_\mu
\nonumber\\
&=&e(-{\cal M}_t(p_a-q,p_b,p_1,p_2)+{\cal M}_t(p_a,p_b-p_a,p_1,p_2)
+{\cal M}_t(p_a,p_b,p_1+q,p_2)-{\cal M}_t(p_a,p_b,p_1,p_2+q)
\nonumber\\
&=&4eg^2_{\sigma\pi\pi}\left\{F^2[(p_b-p_2)^2]
\left[\frac{(p_a\cdot p_1-q\cdot p_1)p_b\cdot p_2}
{(p_b-p_2)^2-m^2_\sigma+im_\sigma\Gamma_\sigma}
-\frac{(p_a\cdot p_1+p_a\cdot q)p_b\cdot p_2}
{(p_b-p_2)^2-m^2_\sigma+im_\sigma\Gamma_\sigma}\right]
\right.\nonumber\\
&&+\left.F[(p_a-p_1)^2]\left[-\frac{p_a\cdot p_1(p_b\cdot p_2-q\cdot p_2)}
{(p_a-p_1)^2-m^2_\sigma+im_\sigma\Gamma_\sigma}
+\frac{p_a\cdot p_1(p_b\cdot p_2+p_b\cdot q)}
{(p_a-p_1)^2-m^2_\sigma+im_\sigma\Gamma_\sigma}\right]\right\}\nonumber\\
&=&4eg^2_{\sigma\pi\pi}\left[F^2[(p_b-p_2)^2]\frac{-(q\cdot p_1+q\cdot p_a)p_b\cdot p_2}
{(p_b-p_2)^2-m^2_\sigma+im_\sigma\Gamma_\sigma}
+F^2[(p_a-p_1)^2]\frac{p_a\cdot p_1(q\cdot p_2+q\cdot p_b)}
{(p_a-p_1)^2-m^2_\sigma+im_\sigma\Gamma_\sigma}\right] \ ,
\end{eqnarray}
so that the pertinent contact term is deduced as
\begin{eqnarray}
{\cal M}^{(c)\mu}_t\cdot q_\mu&=&F^2[(p_b-p_2)^2]
\frac{4eg^2_{\sigma\pi\pi}(p_a\cdot q+p_1\cdot q) p_b\cdot p_2}
{(p_b-p_2)^2-m^2_{\sigma}+im_\sigma\Gamma_\sigma}
-F^2[(p_a-p_1)^2]\frac{4eg^2_{\sigma\pi\pi}
p_a\cdot p_1(p_b\cdot q+p_2\cdot q)}
{(p_a-p_1)^2-m^2_{\sigma}+im_\sigma\Gamma_\sigma} , \
\end{eqnarray}
resulting in
${\cal M}^\mu_t\cdot q_\mu+{\cal M}^{(c)\mu}_t\cdot q_\mu=0$.
Similarly, for the $s$-channel, we have
\begin{eqnarray}
{\cal M}_s^\mu\cdot q_\mu&=&e(J^\mu_a{\cal M}_s(p_a-q,p_b,p_1,p_2)
+J^\mu_b{\cal M}_s(p_a,p_b-q,p_1,p_2)\nonumber\\
&&+J^\mu_1{\cal M}_s(p_a,p_b,p_1+q,p_2)+J^\mu_2{\cal M}_s(p_a,p_b,p_1,p_2+q))\cdot q_\mu\nonumber\\
&=&e(-{\cal M}_s(p_a-q,p_b,p_1,p_2)+{\cal M}_s(p_a,p_b-p_a,p_1,p_2)
+{\cal M}_s(p_a,p_b,p_1+q,p_2)-{\cal M}_s(p_a,p_b,p_1,p_2+q)\nonumber\\
&=&4eg^2_{\sigma\pi\pi}\left[\frac{(p_a\cdot p_b-q\cdot p_b)p_1\cdot p_2}
{(p_1+p_2)^2-m^2_\sigma+im_\sigma\Gamma_\sigma}
-\frac{(p_a\cdot p_b-p_a\cdot q)p_1\cdot p_2}
{(p_1+p_2)^2-m^2_\sigma+im_\sigma\Gamma_\sigma}\right.\nonumber\\
&&\left.-\frac{p_a\cdot p_b(p_1\cdot p_2+q\cdot p_2)}
{(p_a+p_b)^2-m^2_\sigma+im_\sigma\Gamma_\sigma}
+\frac{p_a\cdot p_b(p_1\cdot p_2+p_1\cdot q)}
{(p_a+p_b)^2-m^2_\sigma+im_\sigma\Gamma_\sigma}\right]\nonumber\\
&=&4eg^2_{\sigma\pi\pi}\left[\frac{-(q\cdot p_b-q\cdot p_a)p_1\cdot p_2}
{(p_1+p_2)^2-m^2_\sigma+im_\sigma\Gamma_\sigma}
+\frac{p_a\cdot p_b(q\cdot p_1-q\cdot p_2)}
{(p_a+p_b)^2-m^2_\sigma+im_\sigma\Gamma_\sigma}\right]
\end{eqnarray}
and the required contact term follows as
\begin{eqnarray}
{\cal M}_s^{(c)\mu}\cdot q_\mu
&=&\frac{4eg^2_{\sigma\pi\pi}(p_b\cdot q-p_a\cdot q)
 p_1\cdot p_2}
{(p_1+p_2)^2-m^2_{\sigma}+im_\sigma\Gamma_\sigma}
-\frac{4eg^2_{\sigma\pi\pi}p_a\cdot p_b(p_1\cdot q-p_2\cdot q)}
{(p_a+p_b)^2-m^2_{\sigma}+im_\sigma\Gamma_\sigma}.
\end{eqnarray}
Again, we arrive at
${\cal M}^\mu_s\cdot q_\mu+{\cal M}^{(c)\mu}_s\cdot q_\mu=0$.
It is straightforward to test the rho-meson exchange channels along
similar lines.

%%%%%%%%%%%%%%%%%%%%%%%%%%%%%%%%%%%%%%%%%%%%%
\section{Rate Parameterizations}
\label{app_param}
%%%%%%%%%%%%%%%%%%%%%%%%%%%%%%%%%%%%%%%%%%%%%
The thermal production rate of photons from Bremsstrahlung off
$\pi$-$\pi$ and $\pi$-$K$ scattering as employed in Sec.~\ref{sec_spec}
can be conveniently parametrized as
\begin{eqnarray}
q_0\frac{dR_{\pi\pi}}{d^3q}&=&\frac{-0.00026421+\frac{0.0075394}{0.13007\sqrt{\pi/2}}
     {\rm exp}\left[\frac{-2(T-0.28351)^2}{0.13007^2}\right]}{0.01277}\nonumber\\
     &&\times\frac{-0.00018184+\frac{0.00073983}{\pi[4(q_0-0.054587)^2+0.099656^2]}}
     {[1.+104(0.18-T)^{1.4}(q_0-0.1)]},\nonumber\\
q_0\frac{dR_{\pi K}}{d^3q}&=&\frac{-0.0000411907+\frac{0.0013114}{0.12037\sqrt{\pi/2}}
     {\rm exp}\left[\frac{-2(T-0.27292)^2}{0.12037^2}\right]}{0.00227}\nonumber\\
     &&\times\frac{-0.000021454+\frac{0.0001459}{\pi[4(q_0-0.043159)^2+0.085668^2]}}
     {[1.+605(0.18-T)^{1.7}(q_0-0.1)]}
\end{eqnarray}
These expressions reproduce the exact rates within $\sim$3\% in the
energy interval between 0.1~GeV and 0.5~GeV and in the temperature
range 0.1-0.18~GeV. Pion and kaon chemical potentials are included
as $\mu_\pi(T)$ and $\mu_K(T)$ according to the fireball evolution
employed in this work for central $Pb$(158~AGeV)-$Pb$ collisions,
cf.~eq.~(\ref{chem}). Temperature $T$ and energy $q_0$ are in units
of GeV, and the resulting rates are in units of 1/GeV$^2$fm$^4$.
For more general use, we also provide parameterizations which
explicitly resolve the $q_0$, $T$ and $\mu_\pi$ dependencies:
\begin{eqnarray}
q_0\frac{dR_{\pi\pi}}{d^3q}&=&\frac{[1+q_0/(0.011+0.56T-0.565T^2)]^{-4}
     (0.92+0.17e^{\mu_\pi/0.0247})}
     {3(0.87+519e^{-T/0.026})(0.92+372e^{-T/0.026})} \ ,
\\
q_0\frac{dR_{\pi K}}{d^3q}&=&\frac{[1+q_0/(-0.024+0.773T)]^{-4}
     (0.8+0.223e^{\mu_K/0.05314})\sqrt{0.92+0.17e^{\mu_\pi/0.0247}}}
     {(14.87+3856e^{-T/0.032})\sqrt{0.92+372e^{-T/0.026}}
     (0.89+143e^{-T/0.026})}
\end{eqnarray}
These expressions reproduce the exact rates within $\sim$5\% in the
energy interval between 0.1~GeV and 0.5~GeV and in the temperature
range 0.1-0.18~GeV.

%%%%%%%%%%%%%%%%%%%%%%%%%%%%%%%%%%%%%%%%%%%%%%%%%%%%%%%%%%%%%%%%%%%%%%%


\begin{thebibliography}{99}

\bibitem{Alam:1999sc}
J.~Alam, S.~Sarkar, P.~Roy, T.~Hatsuda and B.~Sinha,
%``Thermal photons and lepton pairs from quark gluon plasma and hot hadronic
%matter,''
Annals Phys. {\bf 286}, 159 (2001).
%  [arXiv:hep-ph/9909267].
%%CITATION = HEP-PH 9909267;%%

\bibitem{peitzmann}
T.~Peitzmann and M.H.~Thoma, Phys. Rep. {\bf 364}, 175 (2002).

\bibitem{Arleo:2004gn}
  F.~Arleo {\it et al.},
  %``Photon physics in heavy ion collisions at the LHC,''
  arXiv:hep-ph/0311131.
  %%CITATION = HEP-PH/0311131;%%

\bibitem{rapp04}
R.~Rapp, Mod. Phys. Lett. {\bf A19}, 1717 (2004).

\bibitem{Stankus:2005eq}
  P.~Stankus,
  %``Direct photon production in relativistic heavy-ion collisions,''
  Ann.\ Rev.\ Nucl.\ Part.\ Sci.\  {\bf 55}, 517 (2005).
  %%CITATION = ARNUA,55,517;%%

\bibitem{kapusta91}
J. Kapusta, P. Lichard, and D. Seibert, Phys. Rev. D {\bf 44}, 2774(1991).

\bibitem{song93}
C. Song, Phys. Rev. C {\bf 47}, 2861 (1993).

\bibitem{Golov93}
V.V.~Goloviznin and K. Redlich, Phys. Lett. {\bf B319}, 520 (1993).

\bibitem{roy96}
P.K.~Roy, D. Pal, S. Sarkar, D.K. Srivastava and B. Sinha,
Phys. Rev. C {\bf 53}, 2364 (1996).

\bibitem{alam03}
J. Alam, P. Roy, and S. Sarkar,  Phys. Rev. C {\bf 68}, 031901 (2003)

\bibitem{TRG04}
S. Turbide, R. Rapp, and C. Gale, Phys. Rev. C {\bf 69}, 014903 (2004).

\bibitem{Haglin:2003sh}
  K.~L.~Haglin,
  %``Rate of photon production from hot hadronic matter,''
  J.\ Phys.\ G {\bf 30}, L27 (2004).
%  [arXiv:hep-ph/0308084].
  %%CITATION = JPHGB,G30,L27;%%

\bibitem{Sriva05}
D.K.~Srivastava, Phys. Rev. C {\bf 71}, 034905 (2005).

\bibitem{wa98-00}
WA98 Collaboration (M. M. Aggarwal {\it et al.}),
Phys. Rev. Lett. {\bf 85}, 3595 (2000).

\bibitem{wa98-04}
WA98 Collaboration (M. M. Aggarwal {\it et al.}),
Phys. Rev. Lett. {\bf 93}, 022301 (2004).


\bibitem{Tur04}
S.~Turbide, private communication (2004).

\bibitem{Rapp:2005fy}
R.~Rapp,
%``Electromagnetic radiation and in-medium effects,''
arXiv:nucl-th/0502020.
%%CITATION = NUCL-TH 0502020;%%

\bibitem{ruckl}
R. R\"uckl, Phys. Lett. {\bf B64}, 39 (1976).

\bibitem{eggers}
H.C.~Eggers, R.~Tabti, C.~Gale, and K.~Haglin,
Phys. Rev. D {\bf 53}, 4822 (1996).

\bibitem{Cleymans93}
J. Cleymans, V. V. Goloviznin, and K. Redlich, Phys. Rev. D {\bf 47},
173 (1993).

\bibitem{Knoll93}
J. Knoll and R. Lenk, Nucl. Phys. {\bf A561}, 501(1993).

\bibitem{haglin93}
K. Haglin, C. Gale, and V. Emel'yanov, Phys. Rev. D {\bf 47}, 973 (1993).

\bibitem{changhui}
C.H.~Li and C.M.~Ko, Nucl. Phys. {\bf A712}, 110 (2002).

\bibitem{donoghue}
J. F. Donoghue, C. Ramirez, and G. Valencia, Phys. Rev. D {\bf 39},
1947 (1989).

\bibitem{pipi-data}
V. Srinivasan {\it et al.}, Phys. Rev. D {\bf 12}, 681 (1975); S. D. Protopoescu
{\it et al.} Phys. Rev. D {\bf 7}, 1279 (1973).

\bibitem{firestone}
A. Firestone, G. Goldhaber, D. Lissauer, and G. H. Trilling, Phys. Rev.
 D {\bf 5}, 2188 (1972).

\bibitem{Rapp:2002}
R. Rapp, Phys. Rev. C {\bf 66}, 017901 (2002).

\bibitem{RW99}
R. Rapp and J. Wambach, Eur. Phys. J. {\bf A6}, 415 (1999).

\bibitem{lpm}
L. Landau and I. Pomeranchuk, Dokl. Akad. Nauk SSSR {\bf 92}, 535 (1953);
{\bf 92}, 735 (1953);
A. B. Migdal, Dokl. Akad. Nauk SSSR {\bf 96}, 49 (1954);
Phys. Rev. {\bf 103}, 1811 (1956).

\bibitem{Rapp:1993ih}
R.~Rapp and J.~Wambach,
%``Selfconsistent description of a thermal pion gas,''
Phys. Lett. {\bf B315}, 220 (1993).
%[arXiv:nucl-th/9306016].
%%CITATION = NUCL-TH 9306016;%%

\bibitem{Rapp:1993bi}
R.~Rapp and J.~Wambach,
%``Pion properties in a hot pi N Delta gas,''
Nucl. Phys. {\bf A573}, 626 (1994).
%[arXiv:nucl-th/9311018].
%%CITATION = NUCL-TH 9311018;%%

\bibitem{Agakichiev:2005ai}
G.~Agakichiev {\it et al.}  [CERES Collaboration],
%``e+ e- pair production in Pb Au collisions at 158-GeV per nucleon,''
Eur. Phys. J. C {\bf 41}, 475 (2005).
%[arXiv:nucl-ex/0506002].
  %%CITATION = NUCL-EX 0506002;%%

\bibitem{Arnaldi:2006jq}
  R.~Arnaldi {\it et al.}  [NA60 Collaboration],
  %``First measurement of the rho spectral function in high-energy nuclear
  %collisions,''
  Phys.\ Rev.\ Lett.\  {\bf 96}, 162302 (2006).
%  [arXiv:nucl-ex/0605007].
  %%CITATION = PRLTA,96,162302;%%

\bibitem{vanHees:2006ng}
H.~van Hees and R.~Rapp,
%``Comprehensive interpretation of thermal dileptons at the SPS,''
Phys. Rev. Lett. {\bf 97}, 102301 (2006).
%%CITATION = HEP-PH 0603084;%%

\bibitem{Grion:2005hu}
N.~Grion {\it et al.}  [CHAOS Collaboration],
%``The pi $\to$ pi pi process in nuclei and the restoration of chiral
%symmetry,''
Nucl. Phys. {\bf A763}, 80 (2005).
%[arXiv:nucl-ex/0508028].
%%CITATION = NUCL-EX 0508028;%%

\bibitem{Starostin:2000cb}
  A.~Starostin {\it et al.}  [Crystal Ball Collaboration],
  %``Measurement of pi0 pi0 production in the nuclear medium by pi- interactions
  %at 0.408-GeV/c,''
  Phys.\ Rev.\ Lett.\  {\bf 85}, 5539 (2000).
  %%CITATION = PRLTA,85,5539;%%

\bibitem{Hats99}
T.~Hatsuda, T.~Kunihiro, and H.~Shimizu,
Phys. Rev. Lett. {\bf 82}, 2840 (1999).

\bibitem{Cabrera:2005wz}
D.~Cabrera, E.~Oset and M.J.~Vicente Vacas,
%``Evaluation of the pi pi scattering amplitude in the sigma-channel at
%finite density,''
Phys. Rev. C {\bf 72}, 025207 (2005).
%[arXiv:nucl-th/0503014].
%%CITATION = NUCL-TH 0503014;%%

\bibitem{Roder:2005qy}
D.~R\"oder,
%``Selfconsistent calculations of sigma-meson properties at finite
%temperature,''
arXiv:hep-ph/0509232.
%%CITATION = HEP-PH 0509232;%%

\bibitem{Halasz:1997xc}
M.A.~Halasz, J.V.~Steele, G.Q.~Li and G.E.~Brown,
%``Photon rates for heavy-ion collisions from hidden local symmetry,''
Phys. Rev. C {\bf 58}, 365 (1998).
%[arXiv:nucl-th/9712006].
%%CITATION = NUCL-TH 9712006;%%

\bibitem{Song:1998um}
C.s.~Song and G.~I.~Fai,
%``Medium effect on photon production in ultrarelativistic nuclear
%collisions,''
Phys. Rev. C {\bf 58}, 1689 (1998).
%[arXiv:nucl-th/9802068].
%%CITATION = NUCL-TH 9802068;%%

\bibitem{Alam:2001ar}
J.e.~Alam, P.~Roy, S.~Sarkar and B.~Sinha,
%``Spectral red-shift versus broadening from photon and dilepton spectra,''
Phys. Rev. C {\bf 67}, 054901 (2003).
%[arXiv:nucl-th/0106038].
%%CITATION = NUCL-TH 0106038;%%

\bibitem{AMY01}
P.~Arnold, G.D.~Moore and L.G.~Yaffe, JHEP 0112 (2001) 009.

\bibitem{Rapp:1997ei}
R.~Rapp, M.~Urban, M.~Buballa and J.~Wambach,
%``A microscopic calculation of photoabsorption cross sections on protons
% and nuclei,''
Phys. Lett. {\bf B417}, 1 (1998).
%[arXiv:nucl-th/9709008].
%%CITATION = NUCL-TH 9709008;%%

\bibitem{bjorken}
J.~Bjorken and L.~McLerran, Phys. Rev. D {\bf 31}, 63 (1985).

\bibitem{Ko86}
C.M.~Ko and C.Y.~Wong, Phys. Rev. C {\bf 33}, 153 (1986).

\bibitem{na49}
NA49 Collaboration (J. B\"achler {\it et al.}), Nucl. Phys. {\bf A661},
45c (1999).

%\bibitem{RCW97}
%R. Rapp, G. Chanfray, and J. Wambach, Nucl. Phys. {\bf A617}, 472
%(1997).


\end{thebibliography}
\end{document}